\documentclass[runningheads,a4paper]{llncs}

\usepackage{amsfonts}
\usepackage{amsmath}
\usepackage[UKenglish]{babel}
\usepackage{paralist}
\usepackage[usenames,dvipsnames,svgnames,x11names,table]{xcolor}\definecolor{LavanderPosta}{HTML}{ebd9fc}
\usepackage{soul}
\usepackage{url}
\usepackage{graphicx}\graphicspath{ {images/} }
\usepackage{wrapfig}
\usepackage{subcaption}             % Support for subfigures
\usepackage{multirow}
\usepackage{mathabx}
\usepackage[export]{adjustbox}
\usepackage[vlined,noend,linesnumbered]{algorithm2e}
\usepackage{float}
\let\oldnl\nl% Store \nl in \oldnl
  \newcommand\nonl{%
    \renewcommand{\nl}{\let\nl\oldnl}}% Remove line number for one line
\usepackage{mathtools}
\usepackage{stmaryrd}
\usepackage{centernot}% \renewcommand*\rmdefault{cmr}
\usepackage{xspace}
\usepackage{todonotes}
\usepackage{braket}
\usepackage{paralist}

\usepackage{scalerel}

\usepackage{makecell}

\usepackage{enumitem}
\newlist{todolist}{itemize}{2}
\setlist[todolist]{label=\huge$\square$}
\usepackage{pifont}
\usepackage{hyperref}
\hypersetup{
  colorlinks = true,
  linkcolor = purple,
  urlcolor  = purple,
  citecolor = teal,
  anchorcolor = teal
}
\usepackage[capitalise]{cleveref}
\crefname{section}{Sec.}{Secs.}

\usepackage{listings}

\definecolor{codegreen}{rgb}{0,0.6,0}
\definecolor{codegray}{rgb}{0.5,0.5,0.5}
\definecolor{codepurple}{rgb}{0.58,0,0.82}
\definecolor{backcolour}{rgb}{0.95,0.95,0.92}

\lstdefinestyle{qosfsa}{
    backgroundcolor=\color{backcolour},   
    commentstyle=\color{codegreen},
    keywordstyle=\color{magenta},
    numberstyle=\tiny\color{codegray},
    stringstyle=\color{codepurple},
    basicstyle=\ttfamily\scriptsize,
    breakatwhitespace=false,         
    breaklines=true,                 
    captionpos=b,                    
    keepspaces=true,                 
    numbers=left,                    
    numbersep=5pt,                  
    showspaces=false,                
    showstringspaces=false,
    showtabs=false,                  
    tabsize=2,
    morekeywords={fsa, qos_attributes, qos_specifications, final_states}
}

\lstdefinestyle{ql}{
    backgroundcolor=\color{backcolour},   
    commentstyle=\color{codegreen},
    keywordstyle=\color{magenta},
    numberstyle=\tiny\color{codegray},
    stringstyle=\color{codepurple},
    basicstyle=\ttfamily\scriptsize,
    breakatwhitespace=false,         
    breaklines=true,                 
    captionpos=b,                    
    keepspaces=true,                 
    numbers=left,                    
    numbersep=5pt,                  
    showspaces=false,                
    showstringspaces=false,
    showtabs=false,                  
    tabsize=2,
    morekeywords={True, Until, qos}
}

\usepackage{xargs}
\newif\ifemi
\newcommand{\aGcomm}[2][check]{%
  \ifthenelse{\equal{#1}{new}}{{\color{red}#2}}{%
         \ifthenelse{\equal{#1}{changed}}{{\color{teal}{#2}}}{%
                \todo[color=orange!20]{\tiny Agus: \color{NavyBlue}#1}%
                {\color{OliveGreen}{#2}}%
         }%
  }%
}
\newcommand{\aGchange}[2]{%
  \ifthenelse{\equal{#2}{removed}}{%
	 \todo[color=orange!20]{\tiny Agus: removed\\\color{NavyBlue}#1}{}
  }{%
	 \todo[color=orange!20]{\tiny Agus: it was\\
		\color{NavyBlue}#1}{%
		\color{OrangeRed}{#2}
	 }%
  }%
}
\newcommand{\eMcomm}[2][check]{%
  \ifthenelse{\equal{#1}{new}}{{\color{red}#2}}{%
         \ifthenelse{\equal{#1}{changed}}{{\color{teal}{#2}}}{%
                \todo[color=orange!20]{\tiny eM: \color{NavyBlue}#1}%
                {\color{OliveGreen}{#2}}%
         }%
  }%
}
\newcommand{\eMchange}[2]{%
  \ifthenelse{\equal{#2}{removed}}{%
	 \todo[color=orange!20]{\tiny eM: removed\\\color{NavyBlue}#1}{}
  }{%
	 \todo[color=orange!20]{\tiny eM: it was\\
		\color{NavyBlue}#1}{%
		\color{OrangeRed}{#2}
	 }%
  }%
}

\DeclareGraphicsExtensions{%
  .png,.PNG,%
  .pdf,.PDF,%
  .jpg,.mps,.jpeg,.jbig2,.jb2,.JPG,.JPEG,.JBIG2,.JB2}

\usepackage[UKenglish]{babel}
  
\usepackage{fixme}
\fxusetheme{color}
\FXRegisterAuthor{eM}{aeM}{\color{orange} {\underline{eM}}}

\usepackage[normalem]{ulem} % underline command breaks over line ends

\usepackage{xifthen}        % for conditional commands
\newcommand{\ifempty}[3]{%
  \ifthenelse{\isempty{#1}}{#2}{#3}%
}

\newcommand{\mkfun}[4][\colorFun]{
  \newcommand{#2}[1][#4]{
    {#1\textsf{#3}}
    \ifempty{##1}{}{
      ({##1})}
  }
}

\newcommand{\mkuop}[4][\colorFun]{
  \newcommand{#2}[1][#4]{
    {#1\textsf{#3}}
    \ifempty{##1}{}{
      \, {##1}}
  }
}

\newcommand{\hidden}[1]{}
\newcommand{\hide}[1]{\eMcomm[\%\%\%]{}}

\newcommand{\cf}[2]{
  \fontsize{#1}{#1}{\selectfont{#2}}
}
\ifemi
\usepackage{showlabels}

\newcommand{\emi}[2]{
  \marginpar{\fcolorbox{red}{shadecolor}{\cf{#1}{{#2}}}}
}
\newcommand{\emic}[2]{\par
  \fcolorbox{red}{shadecolor}{\parbox{\linewidth}{ 
      \color{gray}
      \begin{description}
      \item[{\color{blue} #2}]{\sf #1}
      \end{description}}}
}
\else
\newcommand{\emi}[2]{}
\newcommand{\emic}[2]{}{}
\fi

\newcommand{\sst}{\;\big|\;}
\newcommand{\conf}[1]{\ensuremath{\langle {#1} \rangle}}
\newcommand{\possib}[1]{\ensuremath{\langle {#1} \rangle}}

\newcommand{\bnfdef}{\ ::=\ }
\newcommand{\bnfmid}{\;\ \big|\ \;}

\newcommand{\qand}[1][and]{\quad\text{#1}\quad}

{\bfseries}{\rmfamily}
{\bfseries}{\rmfamily}

\newcommand{\squo}[1]{\lq {#1}\rq}
\newcommand{\quo}[1]{\lq\lq {#1}\rq\rq}
\def\finex{{\unskip\nobreak\hfil
\penalty50\hskip1em\null\nobreak\hfil$\diamond$
\parfillskip=0pt\finalhyphendemerits=0\endgraf}}

\definecolor{shadecolor}{rgb}{1,0.99,0.9}
\definecolor{bg}{rgb}{0.95,0.95,0.95}

\usepackage{mfirstuc}
\usepackage{xstring}

\newcommand{\gsubs}[2]{^{#1} / _{#2}}
\newcommandx{\gsubst}[3][1=\aM,2=q,3=q',usedefault=@]{
  \left \{\gsubs{#3}{#2} \right \}#1
}
\newcommandx{\gsubsts}[5][1=\aM,2=q,3=q',4=q,5=q',usedefault=@]{
  \left \{\gsubs{#3}{#2}, \gsubs{#5}{#4} \right \}#1
}

\newcommandx{\wellpar}[2][1={\aG},2={\aG'},usedefault=@]{\mathit{wf}({#1}, {#2})}

\newif\ifcp
\cpfalse
\newcommand{\gname}[1][i]{\ifcp{\colorNode{\scriptstyle\textsf{#1}}}\else{}\fi}

\newif\ifguard
\guardfalse
\newcommand{\aguard}{\ifguard{\colorGuard \phi}\else{}\fi}

\def\colorGuard{\color{cyan}}
\def\colorPtp{\color{blue}}
\def\colorFun{\color{Navy}}

\def\colorOp{\color{OliveGreen}}
\def\colorNode{\color{LightCoral}}
\def\colorR{\color{OliveGreen}}
\def\colorE{\color{orange}}

\def\colorMsg{\color{BrickRed}}

\newcommand{\fillcolor}{orange!5}

\newcommand{\toolidcol}{teal}%{red!10!blue!90}
\newcommand{\toolid}[1]{\textsf{\textcolor{\toolidcol}{#1}}\xspace}

\newcommand{\chorgram}{\toolid{ChorGram}}

\newcommand{\msg}[1][m]{\mathsf{\colorMsg{#1}}}
\newcommand{\msgset}{\mathcal{\colorMsg{M}}}

\newcommand{\ptp}[1][A]{\ensuremath{\mathsf{\colorPtp{\capitalisewords{#1}}}}}

\newcommand{\p}{\ptp}
\newcommand{\q}{{\ptp[B]}}
\newcommand{\s}{{\ptp[S]}}

\newcommand{\sndint}[1][{\gint[]}]{\mathrm{snd}(#1)}
\newcommand{\rcvint}[1][{\gint[]}]{\mathrm{rcv}(#1)}
\newcommandx{\ggcommon}[3][1=\ptp,2={\aH},3={\aH'},usedefault=@]{f_{#1}}
\newcommandx{\opair}[2][1={\ae},2={\ae'},usedefault=@]{\conf{{#1},{#2}}}
\newcommandx{\hopair}[2][1={\aE},2={\aE'},usedefault=@]{\llparenthesis\, {#1},{#2}\, \rrparenthesis}
\newcommandx{\wf}[2][1={\aG},2={\aG'},usedefault=@]{wf({#1}, {#2})}
\newcommandx{\wb}[2][1={\aG},2={\aG'},usedefault=@]{wb({#1}, {#2})}
\newcommandx{\ws}[2][1={\aG},2={\aG'},usedefault=@]{ws({#1}, {#2})}

\newcommandx{\widx}[2][1={\aW},2={i},usedefault=@]{{#1}[{#2}]}
\newcommandx{\outop}[2][1=\gname,2={}]{{\colorOp{!}}^{{#1}{#2}}}
\newcommandx{\inop}[2][1=\gname,2={}]{{\colorOp{?}}^{{#1}{#2}}}
\newcommandx{\aout}[4][1=a,2=b,3=m,4={},usedefault=@]{
  \achan[#1][#2] \outop[{}] {\msg[#3]} {#4}
}
\newcommandx{\ain}[4][1={\p},2={\q},3=m,4={},usedefault=@]{
  \achan[#1][#2] \inop[{}] {\msg[{#3}]}{#4}
}
\newcommandx{\adep}[1][1={}]{
  \conf{ \aout[@][@][@][@][{#1}], \ain[@][@][@][@][{#1}]}
}

\newcommandx{\hproj}[2][1=\aH, 2=\ptp, usedefault=@]{
  \ifempty{#1}{}{{#1}}\ifempty{#2}{}{{^{\scriptscriptstyle @{#2}}}}
}
\newcommandx{\eproj}[2][1=\aE,2=A, usedefault=@]{
  {{#1}}\ifempty{#2}{}{{^{\scriptscriptstyle @{{\ptp[#2]}}}}}
}

\tikzset{
  component/.style={
    draw,
    fill = white,
    minimum width = 1.5cm,
    minimum height = .5cm,
    drop shadow
  }
}
\tikzset{
  file/.style={
    thin,
    fill = blue!5,
    font = \tt\scriptsize,
    text width = .8cm,
    minimum width = 1.0cm,
    minimum height = .5cm,
    drop shadow
  }
}

\tikzset{
  dataflow/.style={
    thick,
    draw, ->, >=latex,
    dashed,
    OliveGreen
  }
}

\tikzset{
  pipeline/.style={
    thick,
    draw, ->, >=latex,
    double,
    red
  }
}

\tikzset{
  pomsetcloud/.style={
    cloud,
	 cloud puffs=20,
	 cloud ignores aspect,
	 minimum height=.1cm,
	 minimum width=2cm,
	 fill=blue!10,
	 opacity=.5,
	 draw
  }
}

\newcommand{\apom}{r}

\newcommand{\aR}[1][R]{{\colorR{#1}}}

\newcommand{\aConf}{s}

\newcommandx{\detM}[1][1=\aCM,usedefault=@]{\Delta({#1})}
\tikzset{
  cnode/.style={
    shape=circle,
    minimum size = 0mm,
    inner sep = 1pt,
    font=\tiny,
    draw
  },
  carrow/.style={
    ->,
    shorten >=1pt,
    >=stealth',
    auto,
    font=\scriptsize,
    draw,
    sloped
  }
}

\newcommandx{\choranimation}[4][1=1,2=1,3=notempty,4=.35,usedefault=@]{
  \begin{overlayarea}{#1\linewidth}{#2\textheight}
    \begin{tikzpicture}[
      node distance=1cm and 2cm,
      scale=#4,
      every node/.style={transform shape},
      font=\large
      ]
      \node [choreo, align=center] (global){Choreography $\aG$ \\ global viewpoint};
      \node [below= of global] (fake) {};
      \node [choreo, align=center, local, below =of fake] (typei) {$\aCM_1$ \\ Local viewpoint$_1$};
      \node [choreo, align=center, local, left =of typei] (type1) {$\aCM_i$ \\ Local viewpoint$_1$};
      \node [choreo, align=center, local, right=of typei] (typen) {$\aCM_n$ \\ Local viewpoint$_n$};
      \node<+-> (synctxt) at (9,0)  {\textcolor{Navy}{\bf Synchrony}};
      \node<.-> (asynctxt) at (9,-4) {\textcolor{Navy}{\bf Asynchrony}};
      \node<.-> [below=of type1] (fake1) {};
      \node<.-> [below=of typei] (fakei) {};
      \node<.-> [below=of typen] (faken) {};
      \path<+-> [bigar] (global) edge[sloped,above] node {\color{OliveGreen}Project} (type1);
      \path<.-> [bigar] (global) edge[sloped,above] node {\color{OliveGreen}Project} (typei);
      \path<.-> [bigar] (global) edge[sloped,above] node {\color{OliveGreen}Project} (typen);
      \path<.-> [elli] (type1) -- (typei);
      \path<.-> [elli] (typei) -- (typen);
      \node<+-> [process, below=of fake1] (proc1) {Component$_1$};
      \node<.-> [process, below=of fakei] (proci) {Component$_i$};
      \node<.-> [process, below=of faken] (procn) {Component$_n$};
      \path<.-> [bigar,->,dashed,gray] (type1) edge[sloped,above] node {Validate} (proc1);
      \path<.-> [bigar,->,dashed,gray] (typei) edge[sloped,above] node {Validate} (proci);
      \path<.-> [bigar,->,dashed,gray] (typen) edge[sloped,above] node {Validate} (procn);
      \path<.-> [elli] (proc1) -- (proci);
      \path<.-> [elli] (proci) -- (procn);
		\ifempty{#3}{}{
      \node<+-> [process, right=of procn,xshift=4cm] (evolve1) {Component'$_1$};
      \node<.-> [process, right=of evolve1] (evolvei) {Component'$_i$};
      \node<.-> [process, right=of evolvei] (evolven) {Component'$_n$};
      \path<.-> [bigar,blue,dotted] (procn) edge node [above] {evolve/replace/compose} (evolve1);      
      \path<.-> [elli] (evolve1) -- (evolvei);
      \path<.-> [elli] (evolvei) -- (evolven);
      \node<+-> [choreo, align=center, local, above=of evolve1, yshift=1.5cm] (t11) {New $\aCM'_1$ \\ Local viewpoint$_1$};
      \node<.-> [choreo, align=center, local, above=of evolvei, yshift=1.5cm] (t1i) {New $\aCM'_i$ \\ Local viewpoint$_i$};
      \node<.-> [choreo, align=center, local, above=of evolven, yshift=1.5cm] (t1n) {New $\aCM'_n$ \\ Local viewpoint$_n$};
      \path<.-> [elli] (t11) -- (t1i);
      \path<.-> [elli] (t1i) -- (t1n);
      \path<.-> [bigar,->,dashed,gray] (evolve1) edge[sloped,above] node {Extract} (t11);
      \path<.-> [bigar,->,dashed,gray] (evolvei) edge[sloped,above] node {Extract} (t1i);
      \path<.-> [bigar,->,dashed,gray] (evolven) edge[sloped,above] node {Extract} (t1n);
      \node<+> [above=of t1i, yshift=1.5cm] (qm) {\Huge \textcolor{red}{ ??? }};
      \node<.-> [above=of t1i] (dummy) {};
      \node<+-> [choreo,align=center,above=of dummy,scale=.85] (global') {New choreography $\aG'$ \\ global viewpoint};
      \path<.-> [bigar,-] (t11) -- (dummy);
      \path<.-> [bigar,->] (t1i) edge node[right,xshift=1em] {\color{OliveGreen}Synthesise} (global');
      \path<.-> [bigar,-] (t1n) -- (dummy);
		}
    \end{tikzpicture}
  \end{overlayarea}
}%%%end choranimation

\newcommandx{\cm}[2][1=\ptp, 2=\aM]{{#2}_{#1}}
\newcommandx{\achan}[2][1=A,2=B,usedefault=@]{{\ptp[#1]}{\,}{\ptp[#2]}}
\newcommand{\ptpset}{\mathcal{\colorPtp{P}}}
\newcommand{\PSet}{\ptpset}

\newcommand{\oact}{\outop[]}
\newcommand{\iact}{\inop[]}
\newcommand{\tset}{\to}

\newcommand{\csconf}[2]{\conf{\vec{#1} \ ; \ \vec{#2}}}

\newcommand{\TRANSS}[1]{{\xRightarrow{\raisebox{-.3ex}[0pt][0pt]{$\scriptstyle #1$} }}}

\newcommand{\trans}[2][{}]{\,\xrightarrow{#2}_{#1}\,}

\newcommandx{\acfsmout}[3][1=A,2=B,3=m,usedefault=@]{\achan[{#1}][{#2}] \oact {\msg[{#3}]}}
\newcommandx{\acfsmin}[3][1=A,2=B,3=m,usedefault=@]{\achan[{#1}][{#2}] \iact {\msg[{#3}]}}
\newcommandx{\fsaout}[2][1={\p},2={},usedefault=@]{
  \ptp[#1] \ \outop[]\ \msg[{#2}]
}
\newcommandx{\fsain}[2][1={\p},2={},usedefault=@]{
  \ptp[#1] \ \inop[]\ \msg[{#2}]
}

\makeatletter
\newcommand{\linenumfontsize}{\@setfontsize{\linenumfontsize}{3pt}{3pt}}
\makeatother
\newcommand{\bracketColor}[1]{\textcolor{cyan!50!blue!80}{#1}}
\lstdefinelanguage{sgc}{
  basicstyle=\footnotesize,
  commentstyle=\color{Gray},
  morecomment=[l]{..},
  morecomment=[n]{[[}{]]},
  keywords=[0]{repeat,branch,break,sel,let,in,with,unless},
  morekeywords=[1]{test},
  keywordstyle=[0]\color{teal}\sffamily,
  keywordstyle=[1]\color{red}\sffamily,
  mathescape=true,
  literate=
  {=}{{$\colorOp{=}$}}1
  {->}{{${\colorOp \xrightarrow{}}\ $}}2
  {=>}{{${\colorOp \Rightarrow{}}$}}2
  {:}{{$\colorOp{\colon}$}}1
  {|}{{$\gparop$}}1
  {;}{{$\gseqop$}}1
  {+}{{$\gchoop$}}1
  {(}{{\bracketColor{(}}}1
  {)}{{\bracketColor{)}}}1
  {\{}{{\bracketColor{\{}}}1
  {\}}{{\bracketColor{\}}}}1
  {[}{{\bracketColor{[}}}1
  {]}{{\bracketColor{]}}}1
  {[[}{{\color{Gray}{[[}}}2
  {]]}{{\color{Gray}{]]}}}2
  {(o)}{{\gempty}}1
}
\lstdefinelanguage{sgc}{
  commentstyle=\color{Gray},
  morecomment=[l]{..},
  morecomment=[n]{[[}{]]},
  keywords=[0]{repeat,break,branch,sel,let,in,with,unless},
  morekeywords=[1]{test},
  keywordstyle=[0]\color{teal}\sffamily,
  keywordstyle=[1]\color{red}\sffamily,
  mathescape=true,
  literate=
  {=}{{$\colorOp{=}$}}1
  {->}{{${\colorOp \xrightarrow{}}\ $}}2
  {=>}{{${\colorOp \Rightarrow{}}$}}2
  {:}{{$\colorOp{\colon}$}}1
  {|}{{$\gparop$}}1
  {;}{{$\gseqop$}}1
  {+}{{$\gchoop$}}1
  {(}{{\bracketColor{(}}}1
  {)}{{\bracketColor{)}}}1
  {\{}{{\bracketColor{\{}}}1
  {\}}{{\bracketColor{\}}}}1
  {[}{{\bracketColor{[}}}1
  {]}{{\bracketColor{]}}}1
  {[[}{{\color{Gray}{[[}}}2
  {]]}{{\color{Gray}{]]}}}2
  {(o)}{{\gempty}}1
}

\newcommand{\aG}{\mathsf{G}}

\newcommand{\gseqop}{{\colorOp ;}\,}
\newcommand{\gparop}{{\colorOp \ |\ }}
\newcommand{\gchoop}{{\colorOp \ +\ }}
\newcommand{\grecop}{{\colorOp *}}

\newcommandx{\nmerge}[2][1={i},2={},usedefault=@]{
  \ifempty{#2}{
    \ifempty{#1}{\mu}{\gname[-{#1}]}
  }{-{#2}}
}

\newcommand{\gselkw}{\textcolor{OliveGreen}{\mathtt{sel}}}
\newcommand{\grepkw}{\textcolor{OliveGreen}{\mathtt{repeat}}}
\newcommand{\gbrkkw}{\textcolor{OliveGreen}{\mathtt{break}}}

\mkfun{\esbj}{sbj}{\ae}

\makeatletter%
\@ifclassloaded{exam-paper}%
  {}%
  {\makeatletter%
    \@ifclassloaded{test}%
    {}%
    \makeatother%
  }
\makeatother%

\newcommandx{\gnode}[2][1=i,2=\gint,usedefault=@]{
  \ifcp{
    \ifempty{#1}{#2}{\gname[#1].\big({#2}\big)}
  }
  \else
  {#2}
  \fi
}

\newcommand{\gempty}{\mathbf{0}}
\newcommandx{\refgint}[3][1=A,2=\msg,3=B,usedefault=@]{
  \ptp[#1] {\colorOp \xdashrightarrow[{}]{\msg[{#2}]}}{
    \renewcommand{\do}[1]{\ptp[##1]}
	 \docsvlist{#3}
  }
}
\newcommandx{\gint}[4][1=i,2=A,3=m,4=B,usedefault=@]{
  \ptp[#2] {\colorOp \xrightarrow{\scriptstyle \gname[#1]}} \ptp[#4] {\colorOp \colon} {\msg[{#3}]}
}
\newcommandx{\gout}[4][1=\gname,2=\ptp,3=m,4={\ptp[C]},usedefault=@]{
  \achan[{#2}][{#4}] {\colorOp {\colorOp{!}}} {\msg[{#3}]}
}
\newcommandx{\gin}[4][1=\gname,2=\ptp,3=m,4={\ptp[C]},usedefault=@]{
  \achan[{#2}][{#4}] {\colorOp {\colorOp{?}}} {\msg[{#3}]}
}
\newcommandx{\gseq}[3][1=\gname,2={\aG},3={\aG'},usedefault=@]{
  \def\ggraph{{#2} \gseqop {#3}}
  \ggraph
}
\newcommand{\ginfix}[4]{
  \def\ggraph{{#2} {#4} {#3}}
  \gnode[#1][\ggraph]
}
\newcommandx{\gpar}[3][1=i,2={\aG},3={\aG'},usedefault=@]{
  \ginfix{#1}{#2}{#3}{\gparop}
}
\newcommandx{\gcho}[3][1=i,2={\aG},3={\aG'},usedefault=@]{
  \ginfix{#1}{#2}{#3}{\gchoop}
}
\newcommandx{\gchov}[3][1=\gname,2={\aG},3={\aG'},usedefault=@]{
  \def\ggraph{\left(
  \begin{array}l
    \ifempty{#1}{{#2} \\ \gchoop \\ {#3}}{\!\!{#2} \\ \gchoop \\ {#3}}
  \end{array}\right)}
  \ifcp\gnode[{$#1$}][\ggraph] \else \ggraph \fi
}
\newcommandx{\grec}[2][1={},2={\aG},usedefault=@]{
  \def\ggraph{{#2}^\grecop}
  \ifempty{#1}{\ggraph}{\gname[{$#1$}][\ggraph]}
}	

\newcommand{\getcentroid}[2]{
    \coordinate (tmpgatecoord) at (0,0);
    \foreach \n [count=\i] in {#1}{
      \path (\n);
      \coordinate (tmpgatecoord) at ($(tmpgatecoord) + (\n)$);
      \coordinate (#2) at ($1/\i*(tmpgatecoord)$);
    }
}

\tikzset{
  hgsem/.style={
    draw,
    node distance=2cm and 1cm,
    transform shape,
    smooth,
    every node/.style = {font=\sffamily\bfseries}
  }
}

\tikzset{
  hgstyle/.style={
    src color={#1},
    tgt color={#1},
    centroid color={#1},
    centroid label={#1},
    centroid name={#1},
    centroid radius={#1},
    centroid ratio={#1},
    xoffset={#1},
    yoffset={#1},
    xsrcoffset={#1},
    ysrcoffset={#1},
    xtgtoffset={#1},
    ytgtoffset={#1},
    font={#1},
    centroid angle={#1},
    centroid tolerance={#1}
  },
  src color/.store in = \hgsrccol,
  tgt color/.store in = \hgtgtcol,
  centroid color/.store in =\hgfillcolor,
  centroid label/.store in =\hglabel,
  centroid name/.store in =\hgname,
  centroid radius/.store in = \hgradius,
  centroid ratio/.store in = \hgratio,
  xoffset/.store in =\hgxoffset,
  yoffset/.store in =\hgyoffset,
  xsrcoffset/.store in =\hgxsrcoffset,
  ysrcoffset/.store in =\hgysrcoffset,
  xtgtoffset/.store in =\hgxtgtoffset,
  ytgtoffset/.store in =\hgytgtoffset,
  centroid angle/.store in =\hgangle,
  centroid tolerance/.store in =\hgtolerance,
  src color = black,
  tgt color = black,
  centroid color = orange!40,
  centroid label={},
  centroid name={dummycentroid},
  centroid radius = .7pt,
  centroid ratio = .35,
  xoffset = 0,
  yoffset = 0,
  xsrcoffset = 0,
  ysrcoffset = 0,
  xtgtoffset = 0,
  ytgtoffset = 0,
  font=\sffamily\scriptsize,
  centroid angle=0,
  centroid tolerance=10pt
}

\newcommandx{\mkhg}[5][1={},4={},5={},usedefault=@]{
  \begingroup
  \tikzset{#1}
  \StrCount{#2,}{,}[\l] % from package xstring
  \StrCount{#3,}{,}[\m] % from package xstring
  \ifthenelse{\l = 1 \AND \m = 1}{
    \ifempty{#4}{
      \ifempty{#5}{
        \path[hgsem, ->, >=stealth', shorten >=1pt] (#2) -- (#3);
      }{
        \path[hgsem, ->, >=stealth', shorten >=1pt] (#2) #5 (#3);
      }
    }{
      \ifempty{#5}{
        \path[hgsem, ->, >=stealth', shorten >=1pt, #4] (#2) -- (#3);
      }{
        \path[hgsem, ->, >=stealth', shorten >=1pt, #4] (#2) #5 (#3);
      }
    }
  }{
    \coordinate (srcoffset) at (\hgxsrcoffset,\hgysrcoffset);
    \coordinate (tgtoffset) at (\hgxtgtoffset,\hgytgtoffset);
    \getcentroid{#2}{srccentroid};
    \getcentroid{#3}{tgtcentroid};
    \node[label={left:\hglabel}] (\hgname) at ($(srccentroid)!{1-\hgratio}!\hgangle:(tgtcentroid) + (\hgxoffset,\hgyoffset)$) {};
    \pgfgetlastxy \xc \yc;
    \pgfmathtruncatemacro{\xcontrol}{\xc};
    \pgfmathtruncatemacro{\ycontrol}{\yc};
    \foreach \n in {#2}{
      \path (\n);
      \pgfgetlastxy \xntmp \yntmp;
      \pgfmathtruncatemacro{\xn}{\xntmp};
      \pgfmathtruncatemacro{\yn}{\yntmp};
      \pgfmathsetmacro\xtmpdiff{abs(\xn - \xcontrol + \hgxsrcoffset)};
      \pgfmathsetmacro\ytmpdiff{abs(\yn - \ycontrol + \hgytgtoffset)};
      \ifdim \xtmpdiff pt > \hgtolerance
      \ifempty{#4}{
        \path[hgsem, \hgsrccol] (\n) .. controls ($(srccentroid.center) + (srcoffset)$) .. (\hgname.center);
      }{
        \path[hgsem, \hgsrccol] (\n) .. controls ($(srccentroid.center) + (srcoffset)$) .. (\hgname.center);
      }
      \else
      \ifempty{#4}{
        \path[hgsem, \hgsrccol] (\n) -- (\hgname.center);
      }{
        \path[hgsem, \hgsrccol, #4] (\n) -- (\hgname.center);
      }
      \fi
    }
    \foreach \n in {#3}{
      \path (\n);
      \pgfgetlastxy \xntmp \yntmp;
      \pgfmathtruncatemacro{\xn}{\xntmp};
      \pgfmathtruncatemacro{\yn}{\yntmp};
      \pgfmathsetmacro\xtmpdiff{abs(\xn - \xcontrol)};
      \pgfmathsetmacro\ytmpdiff{abs(\yn - \ycontrol)};
      \ifdim \xtmpdiff pt > \hgtolerance
      \ifempty{#4}{
        \path[hgsem, ->, >=stealth', shorten >=1pt, \hgtgtcol] (\hgname.center) .. controls (tgtcentroid.center) and ($(tgtcentroid.center) + (tgtoffset)$) .. (\n);
      }{
        \path[hgsem, ->, >=stealth', shorten >=1pt, \hgtgtcol,#4] (\hgname.center) .. controls (tgtcentroid.center) and ($(tgtcentroid.center) + (tgtoffset)$) .. (\n);
      }
      \else
      \ifempty{#4}{
        \path[hgsem, ->, >=stealth', shorten >=1pt, \hgtgtcol] (\hgname.center) --  (\n);
      }{
        \path[hgsem, ->, >=stealth', shorten >=1pt, \hgtgtcol] (\hgname.center) --  (\n);
      }
      \fi
    }
    \fill[\hgfillcolor] (\hgname) circle [radius=\hgradius];
  }
  \endgroup
}

\newcommandx{\hgordeq}[1][1={\aH},usedefault=@]{\sqsubseteq_{#1}}
\newcommandx{\gintsem}[4][4=.5]{
  \tikz[hgsem,scale=#4,every node/.style={font=\scriptsize}]{
    \node (out) {$\aout[{#1}][{#2}][][{#3}]$};
    \node[below = 20pt of out] (in) {$\ain[{#1}][{#2}][][{#3}]$};
    \mkhg{out}{in};
  }
}

\newcommandx{\gsem}[2][1={\aG},2={},usedefault=@]{\left\llbracket {#1} \right\rrbracket_{#2}}

\newcommandx{\rbot}{\text{undef}}
\newcommandx{\rtrs}[1][1={\aH},usedefault=@]{{#1}^{\star}}
\newcommandx{\gord}[1][1={\aG},usedefault=@]{\leq_{#1}}
\newcommandx{\gordeq}[1][1={\aG},usedefault=@]{\leq_{#1}}
\mkfun{\cause}{cs}{}
\mkfun{\effect}{ef}{}
\newcommandx{\aW}{w}
\newcommandx{\rlang}{\mathcal{L}}

\newcommand{\gfun}[1]{\ensuremath{\mathsf{\colorFun #1}}}
\mkfun{\eact}{\gfun{act}}{}
\mkfun{\enode}{\gfun{cp}}{}

\mkuop{\rmax}{\gfun{max}}{\aH}
\mkuop{\rmin}{\gfun{min}}{\aH}
\mkuop{\rMAX}{\gfun{lst}}{\aH}
\mkuop{\rMIN}{\gfun{fst}}{\aH}

\newcommandx{\rseq}[2][1=\aG,2={\aG'},usedefault=@]{\gfun{seq}({#1},{#2})}
\newcommandx{\rpar}[2][1=\aG,2={\aG'},usedefault=@]{\gfun{par}({#1},{#2})}

\newcommandx{\gproj}[2][1=\aG,2=\ptp]{{#1}\downarrow_{#2}}
\newcommandx{\cinit}[1][1={\aQzero},usedefault=@]{{#1}}
\newcommandx{\cfinal}[1][1={q_e},usedefault=@]{{#1}}

\newcommandx{\geproj}[4][1=\aG,2=\ptp,3=\cinit,4=\cfinal,usedefault=@]{
  {#1}\downarrow_{#2}^{{#3},{#4}}
}

\newcommand*{\StrikeThruDistance}{0.15cm}%
\tikzset{strike thru arrow/.style={
    decoration={markings, mark=at position 0.5 with {
        \draw [blue, thick,-] 
            ++ (-\StrikeThruDistance,-\StrikeThruDistance) 
            -- ( \StrikeThruDistance, \StrikeThruDistance);}
    },
    postaction={decorate},
}}

\newcommandx{\ich}[1][1={\aG},usedefault=@]{{#1}^{\oplus}}
\newcommandx{\ichedges}[2][1={\aG},2={\gname},usedefault=@]{{#1}^{\oplus}({#2})}
\newcommandx{\parts}[1]{2^{#1}}
\newcommandx{\actch}{c}
\newcommandx{\soundactch}[2][1={\aG},2={\actch},usedefault=@]{{#1} \,\circledR\, {#2}}
\newcommandx{\rOnActch}[2][1={\aG},2={\actch},usedefault=@]{{#1} \setminus {#2}}
\newcommandx{\rOnActchClean}[2][1={\aG},2={\actch},usedefault=@]{{#1} \circledR {#2}}
\newcommandx{\rAllEvents}[1][1={\aG},usedefault=@]{\mathit{dom}(#1)}

\newcommand{\AV}{\mathcal{V}}
\newcommand{\aH}{H}

\newcommandx{\hgvertex}[2][1=\al,2=\gname,usedefault=@]{{#1}_{\textcolor{red}{[{#2}]}}}
\newcommand{\aE}{{\colorE E}}
\renewcommand{\ae}[1][e]{{\colorE{#1}}}

\newcommand{\al}[1][\ell]{{\colorE{#1}}}
\newcommandx{\hyedge}[1]{\{#1\}}

\newcommandx{\rdiv}[2][1=\gcho,2=\ptp,usedefault=@]{
  \gfun{div}_{#2}(#1)
}

\newcommandx{\rrdiv}[5][1={\aG},2={\aG'},3={\AV},4={,\AV'},5=\ptp,usedefault=@]{
  \gfun{div}^{#3#4}_{#5}(#1,#2)
}
\newcommandx{\pdiv}[3][1={\apom_1},2={\apom_2},3={\apom},usedefault=@]{
  \gfun{div}_{#3}(#1,#2)
}
\newcommandx{\pfork}[3][1={\apom_1},2={\apom_2},3={\apom},usedefault=@]{
  \gfun{fork}_{#3}(#1,#2)
}

\tikzset{
  pomset/.style={
    node distance = .6cm and .6cm,
    scale = .7,
    transform shape,
    smooth
  }
}

\newcommandx{\mkint}[6][3=i,4=\p,5=\msg,6=\q,usedefault=@]{
  \node[bblock, #1] (#2) {$\gint[#3][#4][#5][#6]$};
}

\newcommand{\mkseq}[2]{\path[line] (#1) -- (#2);}

\newcommand{\mknseq}[1]{
  \StrCount{#1}{,}[\l] % from package xxstring
  \StrBefore{#1}{,}[\myhead]
  \StrBehind{#1}{,}[\mytail]
  \StrBefore{\mytail}{,}[\sndel]
  \ifnum \l > 1 {
    \mkseq{\myhead}{\sndel};
    \mknseq{\mytail}
  }
  \else{\ifnum \l > 0{
      \mkseq{\myhead}{\mytail};
    }
    \else{}
    \fi
  }
  \fi
}

\newcommandx{\mkgateblock}[6][6=yellow!10,usedefault=@]{
  \path(#2);
  \pgfgetlastxy{\xgate}{\ygate};
  \pgfmathtruncatemacro{\xgateround}{\xgate};
  \StrCount{#3,}{,}[\l] % from package xstring
  \ifnum \l < 2 {\errmessage{#3 argument should be a comma-separated list of lenght >= 2}}
  \else{
    \foreach \n in {#3}{
      \path (\n);
      \pgfgetlastxy{\xnode}{\ynode};
      \pgfmathtruncatemacro{\xnround}{\xnode};
      \pgfmathsetmacro\tmpdiff{abs(\xnround - \xgateround)}
      \ifdim \tmpdiff pt > 1 pt \path[line] (#2) -| (\n);
      \else
        \path[line] (#2) -- (\n);
      \fi
    }
  }
  \fi
  \StrCount{#4,}{,}[\l] % from package xstring
  \ifnum \l < 2 {\errmessage{#4 argument should be a comma-separated list of lenght >= 2}}
  \else{
    \foreach \n in {#4}{
      \path (\n);
      \pgfgetlastxy{\xnode}{\ynode};
      \pgfmathtruncatemacro{\xnround}{\xnode};
      \pgfmathsetmacro\tmpdiff{abs(\xnround - \xgateround)}
      \ifdim \tmpdiff pt > 1 pt \path[line] (\n) |- (#5);
      \else
        \path[line] (\n) -- (#5);
      \fi
    }
  }
  \fi
  \node[#1] at (#2) {};
  \node[#1] at (#5) {};
  {
    \begin{pgfonlayer}{background}
      \path[fill=#6,rounded corners]
      (current bounding box.south west) rectangle
      (current bounding box.north east);
    \end{pgfonlayer}
  }
}

\newcommandx{\mkbranchblock}[5][5=@]{
  \mkgateblock{ogate}{#1}{#2}{#3}{#4}[#5]
}

\newcommandx{\mkforkblock}[5][5=@]{
  \mkgateblock{agate}{#1}{#2}{#3}{#4}[#5]
}

\newcommandx{\mkgraph}[3][1=.5cm, usedefault=@]{
  \node[source,above = #1 of {#2}] (src#2) {};
  \node[sink,below  = #1 of {#3}] (sink#3) {};
  \path[line] (src#2) -- (#2);
  \path[line] (#3) -- (sink#3);
}

\newcommandx{\mkloop}[5][1=.5, 2=1.5, 5=\aguard, usedefault=@]{
  \node[lgate,above = #1 of {#3}] (entry#3) {};
  \pgfgetlastxy \xentry \yentry;
  \pgfmathtruncatemacro{\xentryrounded}{\xentry};
  \node[below = #1 of {#4}, label = {above right:{$#5$}},yshift=-1em] (dummy) {};
  \node[lgate,below  = #1 of {#4}] (exit#4) {};
  \pgfgetlastxy \xexit \yexit;
  \pgfmathtruncatemacro{\xexitrounded}{\xexit};
  \path[line] (entry#3) -- (#3);
  \path[line] (#4) -- (exit#4);
  \pgfmathsetmacro\tmpdiff{abs(\xentryrounded - \xexitrounded)}
  \path[line, color=teal] (exit#4) -| ($(exit#4)+(\tmpdiff,0)+(#2,0)$) |- (entry#3);
}

\newcommandx{\mkfork}[4][2=gatenode,3=i,4=.6,usedefault=@]{
  \mkgatebegin{#1}[{\gname[{#3}]}][agate][#4]{#2}
}

\newcommandx{\mkbranch}[4][2=gatenode,3=i,4=.6,usedefault=@]{
  \mkgatebegin{#1}[{\gname[{#3}]}][ogate][#4]{#2}
}

\newcommandx{\mkgatebegin}[5][2={},3=ogate,4=.5,usedefault=@]{
  \coordinate (gatecord) at (0,0);
  \coordinate (xmax) at (0,0);
  \coordinate (xmin) at (0,0);
  \pgfgetlastxy \xmin \xmax;
  \foreach \n [count=\i] in {#1}{
    \pgfgetlastxy \xc \yc;
    \path (\n);
    \pgfgetlastxy \xn \yn;
    \ifnum \i = 1
      \coordinate (xmin) at (\xn,0);
      \coordinate (xmax) at (\xn,0);
      \coordinate (max) at (0,\yn);
    \else
      \ifdim \xn < \xmin
        \coordinate (xmin) at (\xn,0);
      \fi
      \ifdim \xn > \xmax
        \coordinate (xmax) at (\xn,0);
      \fi
      \ifdim \yn < \yc
        \coordinate (max) at (0,\yc);
      \else
        \coordinate (max) at (0,\yn);
      \fi
    \fi
  }
  \coordinate (gatecord) at ($(xmin)!.5!(xmax) + (max) + (0,#4) + (max)$);
  \node[#3,label={below:$#2$}] (#5) at (gatecord) {};
  \pgfgetlastxy{\xgate}{\ygate};
  \pgfmathtruncatemacro{\xgateround}{\xgate};
  \StrCount{#1,}{,}[\l] % from package xxstring
  \ifnum \l < 2 {\errmessage{#1 argument should be a comma-separated list of lenght >= 2}}
  \else{
    \foreach \n in {#1}{
      \path (\n);
      \pgfgetlastxy{\xnode}{\ynode};
      \pgfmathtruncatemacro{\xnround}{\xnode};
      \pgfmathsetmacro\tmpdiff{abs(\xnround - \xgateround)}
      \ifdim \tmpdiff pt > 1 pt \path[line] (#5) -| (\n);
      \else
        \path[line] (#5) -- (\n);
      \fi
    }
  }
  \fi
}

\newcommandx{\mkgatebeginold}[5][2={},3=ogate,4=.5,usedefault=@]{
  \coordinate (gatecord) at (0,0);
  \foreach \n [count=\i] in {#1}{
    \pgfgetlastxy \xc \yc;
    \path (\n);
    \pgfgetlastxy \xn \yn;
    \coordinate (gatecord) at ($(gatecord) + (\xn,0)$);
    \coordinate (gatecord) at ($1/\i*(gatecord)$);
    \ifdim \yn < \yc
    \node (max) at (0,\yc) {};
    \else
    \node (max) at (0,\yn) {};
    \fi
  }
  \coordinate (gatecord) at ($(gatecord) + (0,#4) + (max)$);
  \node[#3,label={below:$#2$}] (#5) at (gatecord) {};
  \pgfgetlastxy{\xgate}{\ygate};
  \pgfmathtruncatemacro{\xgateround}{\xgate};
  \StrCount{#1,}{,}[\l] % from package xxstring
  \ifnum \l < 2 {\errmessage{#1 argument should be a comma-separated list of lenght >= 2}}
  \else{
    \foreach \n in {#1}{
      \path (\n);
      \pgfgetlastxy{\xnode}{\ynode};
      \pgfmathtruncatemacro{\xnround}{\xnode};
      \pgfmathsetmacro\tmpdiff{abs(\xnround - \xgateround)}
      \ifdim \tmpdiff pt > 1 pt \path[line] (#5) -| (\n);
      \else
        \path[line] (#5) -- (\n);
      \fi
    }
  }
  \fi
}

\newcommandx{\mkmerge}[4][2=gatenode,3=i,4=.5,usedefault=@]{
  \mkgateend{#1}[{\ifempty{#3}{}{\nmerge[#3]}}][ogate][#4]{#2}
}

\newcommandx{\mkjoin}[4][2=gatenode,3=i,4=.5,usedefault=@]{\mkgateend{#1}[{\ifempty{#3}{}{\nmerge[#3]}}][agate][#4]{#2}}

\newcommandx{\mkgateend}[5][2={},3=ogate,4=.5,usedefault=@]{
  \coordinate (gatecord) at (0,0);
  \coordinate (xmax) at (0,0);
  \coordinate (xmin) at (0,0);
  \pgfgetlastxy \xmin \xmax;
  \foreach \n [count=\i] in {#1}{
    \pgfgetlastxy \xc \yc;
    \path (\n);
    \pgfgetlastxy \xn \yn;
    \ifnum \i = 1
      \coordinate (xmin) at (\xn,0);
      \coordinate (xmax) at (\xn,0);
      \coordinate (ymin) at (0,\yn);
    \else
      \ifdim \xn < \xmin
        \coordinate (xmin) at (\xn,0);
      \fi
      \ifdim \xn > \xmax
        \coordinate (xmax) at (\xn,0);
      \fi
      \ifdim \yn > \yc
        \coordinate (ymin) at (0,\yc);
      \else
        \coordinate (ymin) at (0,\yn);
      \fi
    \fi
  }
  \coordinate (gatecord) at ($(xmin)!.5!(xmax) + (ymin)$);
  \node[#3,label={above:$#2$}] (#5) at ($(gatecord) - (0,{#4})$) {};
  \pgfgetlastxy{\xgate}{\ygate};
  \pgfmathtruncatemacro{\xgateround}{\xgate};
  \StrCount{#1,}{,}[\l] % from package xstring
  \ifnum \l < 2 {\errmessage{#1 argument should be a comma-separated list of lenght >= 2}}
  \else{
    \foreach \n in {#1}{
      \path (\n);
      \pgfgetlastxy{\xnode}{\ynode};
      \pgfmathtruncatemacro{\xnround}{\xnode};
      \pgfmathsetmacro\tmpdiff{abs(\xnround - \xgateround)}
      \ifdim \tmpdiff pt > 1 pt \path[line] (\n) |- (#5);
      \else
        \path[line] (\n) -- (#5);
      \fi
    }
  }
  \fi
}

\newcommandx{\mkgateendold}[5][2={},3=ogate,4=.5,usedefault=@]{
  \coordinate (gatecord) at (0,0);
  \coordinate (xmax) at (0,0);
  \coordinate (xmin) at (0,0);
  \pgfgetlastxy \xmin \xmax;
  \foreach \n [count=\i] in {#1}{
    \pgfgetlastxy \xc \yc;
    \path (\n);
    \pgfgetlastxy \xn \yn;
    \ifdim \xn < \xmin
    \coordinate (xmin) at (\xn,0);
    \fi
    \ifdim \xn > \xmax
    \coordinate (xmax) at (\xn,0);
    \fi
    \ifdim \yn > \yc
    \coordinate (ymin) at (0,\yc);
    \else
    \coordinate (ymin) at (0,\yn);
    \fi
    \coordinate (gatecord) at ($(xmin)!.5!(xmax) + (ymin)$);
  }
  \node[#3,label={above:$#2$}] (#5) at ($(gatecord) - (0,{#4})$) {};
  \pgfgetlastxy{\xgate}{\ygate};
  \pgfmathtruncatemacro{\xgateround}{\xgate};
  \StrCount{#1,}{,}[\l] % from package xstring
  \ifnum \l < 2 {\errmessage{#1 argument should be a comma-separated list of lenght >= 2}}
  \else{
    \foreach \n in {#1}{
      \path (\n);
      \pgfgetlastxy{\xnode}{\ynode};
      \pgfmathtruncatemacro{\xnround}{\xnode};
      \pgfmathsetmacro\tmpdiff{abs(\xnround - \xgateround)}
      \ifdim \tmpdiff pt > 1 pt \path[line] (\n) |- (#5);
      \else
        \path[line] (\n) -- (#5);
      \fi
    }
  }
  \fi
}

\newcommand{\gatedistancein}{3pt}
\newcommand{\gatedistanceinand}{2pt}

\usetikzlibrary{
  arrows,
  backgrounds,
  chains,
  calc,
  decorations.markings,decorations.pathreplacing,
  fadings,
  fit,
  patterns,
  petri,
  positioning,
  shadows,
  shapes,automata,shapes.callouts
}

\tikzset{
  src/.style={draw,circle,fill=white,
    minimum size=2mm,
    inner sep=0pt
  },
  sink/.style={draw,circle,double,fill=white,
    minimum size=1.5mm,
    inner sep=0pt
  },
  node/.style={draw,circle,fill=black,
    minimum size=2mm,
    inner sep=0pt
  },
  source/.style={draw,circle,fill=white,
    minimum size=3mm,
    inner sep=0pt
  },
  sink/.style={draw,circle,double,fill=white,
    minimum size=3mm,
    inner sep=0pt
  },
  block/.style = {rectangle, draw=gray, align=center, fill=orange!25, rounded corners=0.1cm,
    minimum size=5mm, inner sep=2pt},
  prenode/.style = {minimum size=9pt,inner sep=2pt, font=\Large},
  bblock/.style = {rectangle, draw=blue!50, opacity=.7, line width=.5pt, align=center, fill=white, rounded corners=0.1cm,
    minimum size=4mm, inner sep=1pt},
  prenode/.style = {minimum size=9pt,inner sep=2pt, font=\Large},
  agate/.style={draw, rectangle,
    minimum size=3mm,
    inner sep=0pt,
    fill=orange!25,
    label={[red]center:$\mid$}
  },
  ogate/.style = {
    diamond, draw, fill=orange!25,
    minimum size=4mm,
    inner sep=0pt,
    label={[red]center:$+$}
  },
  lgate/.style = {
    diamond, draw, fill=orange!25,
    minimum size=4mm,
    inner sep=0pt,
    label={[red]center:$\circlearrowleft$}
    },
  altogate/.style = {
    diamond, draw,
    minimum size=4mm,
    inner sep=0pt,
    postaction={path picture={% 
        \draw
        ([yshift=\gatedistancein]path picture bounding box.south) -- ([yshift=-\gatedistancein]path picture bounding box.north)
        ([xshift=-\gatedistancein]path picture bounding box.east) -- ([xshift=\gatedistancein]path picture bounding box.west)
        ;}}},
  altgate/.style={draw, rectangle,
    minimum size=3mm,
    inner sep=0pt,
    postaction={path picture={% 
        \draw
        ([yshift=\gatedistanceinand]path picture bounding box.south) --
        ([yshift=-\gatedistanceinand]path picture bounding box.north) ;}}},
  anygate/.style = {circle, draw, fill=white,
    minimum size=4mm,
    inner sep=0pt,
    postaction={path picture={% 
        \draw[black]
        ([xshift=-\gatedistancein,yshift=\gatedistancein]path picture bounding box.south east) --
        ([xshift=\gatedistancein,yshift=-\gatedistancein]path picture bounding box.north west)
        ([xshift=-\gatedistancein,yshift=-\gatedistancein]path picture bounding box.north east) --
        ([xshift=\gatedistancein,yshift=\gatedistancein]path picture bounding box.south west)
        ;}}
  },
  smallglobal/.style={
        node distance=1cm and 0.8cm, semithick, scale=0.8, every node/.style={transform shape}
  },
  elli/.style = {draw,densely dotted,-},
  line/.style = {draw,->, rounded corners=0.07cm,>=latex},
  nline/.style = {draw,semithick, ->},
  pline/.style = {draw,->,>=latex},
  node distance=1cm and 0.7cm,
  baseline=(current  bounding  box.center),
  local/.style={rectangle, draw, fill=\fillcolor, drop shadow,
    text centered, rounded corners, minimum height=5em
  },
  bigar/.style={
    draw,very thick, ->
  },
  process/.style={rectangle, draw=gray, fill=\fillcolor, drop shadow,
    text centered, minimum height=5em,text=gray
  },
  choreo/.style={rectangle, draw, fill=\fillcolor, drop shadow,
    text centered, rounded corners, minimum height=5em
  },
  mycfsm/.style={
        font=\footnotesize,
        initial where=above,
        ->,>=stealth,auto, node distance=1cm and 1cm,
        scale=1, every node/.style={transform shape},
        every state/.style=inner sep=2pt,
        baseline=(current  bounding  box.center),
        initial text={}
  },
  machinecloud/.style={
    cloud, cloud puffs=10, cloud ignores aspect, minimum height=.1cm, minimum width=2cm, draw
  },
  fitting node/.style={
    inner sep=0pt,
    fill=none,
    draw=none,
    reset transform,
    fit={(\pgf@pathminx,\pgf@pathminy) (\pgf@pathmaxx,\pgf@pathmaxy)}
  },
  mypetri/.style={
    font=\footnotesize,
    baseline=(current  bounding  box.center)
  },
  silentrans/.style = {rectangle, draw=black, align=center, fill=black,
    minimum height=1pt,
    minimum width=15pt,
    inner sep=1.5pt
  },
  reset transform/.code={\pgftransformreset},
  tmtape/.style={draw,minimum size=1.2cm}
}

\newcommand{\gunlessop}{\mbox{\colorOp\tiny\tt unless}}

\newcommandx{\gtry}[5][1=\gname,2={\aG_1 \gchoop \cdots \gchoop \aG_n},3=\gin,4=\gout,5={j},usedefault=@]{
  \def\foo{\gtryop\ {#2} \ \gcatchop\ {#3} {\colorOp \Rightarrow} {#4} {\colorOp \bullet} {\gname[{#5}]}}
  \gnode[{#1}][{\ifempty{#1} {\foo } {(\foo)}}]
}

\newcommandx{\gtrycatch}[4][1=\gname,2={\aG},3=\gin,4={\aG'},usedefault=@]{
  \def\foo{\gtryop\ {#2} \ \gcatchop\ {#3} \gdoop\ {#4}}
  \gnode[{#1}][{\ifempty{#1} {\foo} {(\foo)}}]
}

\newcommandx{\agG}[2][1={\aG},2=\aguard]{{#1} \ifempty{#2}{}{\ \gunlessop\ {#2}}}

\newcommandx{\grcho}[5][1=\gname,2={\agG},3={\agG[\aG'][\aguard']},4={\cdots},5=A,usedefault=@]{
  \def\foo{{#2} {\ \ifempty{#4}{\gchoop}{\gchoop \ifempty{#4}{}{\ {#4}\  \gchoop}}\ } {#3}}
  \ifempty{#1}{\ifempty{#5}{\foo}{\gselop\ \cpt[{#1}][{\ptp[#5]}]\big\{ \foo \big\}}}{\gselop\ \cpt[{#1}][{\ptp[#5]}]\big\{ \foo \big\}}
}

\newcommandx{\ggprefix}[3][1=\ptp,2={\aR},3={\aR'},usedefault=@]{f_{#1}} % it was \newcommandx{\common}{...}
\newcommand{\aconfigfn}{\chi}
\newcommand{\aconfig}{\ell}

\newcommand{\lstates}{\statemap}
\newcommandx{\sysconfig}[3][1=\lstates,2=\aconfigfn,3={},usedefault=@]{
  \conf{ {#1},{#2} \ifempty{#3}{}{, #3} }
}
\newcommand{\sysctxfn}[1][]{\gamma_{#1}}
\newcommandx{\sysctx}[2][1=\aQ,2={},usedefault=@]{({#1},\sysctxfn[{#2}])}

\newcommandx{\alog}[4][1=\msg,2=q,3=\gname,4=t,usedefault=@]{({#1},{#2},{#3},{#4})}

\newcommand{\aCM}{M}\newcommand{\aM}{\aCM}
\newcommand{\aQ}{Q}
\newcommandx{\aQzero}[1][1=,usedefault=@]{
  {\ifempty{#1}{q_0}{q_{0#1}}}
}
\newcommand{\badbranches}[1][]{\beta\ifempty{#1}{}{({#1})}}
\newcommand{\aTrs}{\tset}
\newcommandx{\guardedaction}[2][1=\al,2=\aguard,usedefault=@]{
  {#1} \ifempty{#2}{}{/} {#2}
}
\newcommandx{\atrM}[4][1=q,2=\al,3={\hat q,\hat \al, \aguard},4=q',usedefault=@]{
  {#1} \xrightarrow[{#3}]{\guardedaction[{#2}][]} {{#4}}
}
\newcommandx{\atrS}[5][
  1={\sysconfig[@][@][\badbranches]},
  2=\al,
  3=\aguard,
  4={\sysconfig[\lstates'][\aconfigfn'][\badbranches]},
  5=\sysctx,usedefault=@
]{
  {#1} \xRightarrow{\qquad} {{#4}}
}
\newcommandx{\arevtrS}[2][
  1={\sysconfig[@][@][\badbranches]},
  2={\sysconfig[\lstates'][\aconfigfn'][\badbranches']},
  usedefault=@
]{
  {#1} \rightsquigarrow {#2}
}
\newcommand{\aCS}{S}
\newcommand{\abuffer}{\vec b}

\newcommandx{\enables}[2][1=\aconfigfn,2=\aguard,usedefault=@]{{#1} \vdash {#2}}
\newcommandx{\gprojfn}[5][1=\aG,2=\ptp,3=\cinit,4=\cfinal,5={},usedefault=@]{
  \mathbf{proj}_{#2}({#1},{#3},{#4}\ifempty{#5}{}{,{#5}})
}

\newcommandx{\rbp}[3][1=\aG,2=\aconfigfn,3=\achan,usedefault=@]{\mathtt{RBP}_{{#1},{#2}}\ifempty{#3}{}{({#3})}}

\newcommand{\apseudoCFSM}{\mathtt{M}}
\newcommandx{\pseudoseq}[2][1=\apseudoCFSM,2=\apseudoCFSM',usedefault=@]{{#1}  ; {#2}}
\newcommandx{\pseudoCFSM}[4][1=\aQ,2=\aQzero,3=\cfinal,4=\aTrs,usedefault=@]{(#1 \ ; #2 \ ; #3 \ ; #4)}
\newcommandx{\markt}[3][1=\hat{\al},2=\hat{q},3=\aguard,usedefault=@]{\%\big({#1} , {#2}, {#3}\big)}

\newcommandx{\borderfn}[2][1=\aconfig,2=\aloop,usedefault=@]{
  \mathsf{border}_{{#2}}\ifempty{#1}{}{({#1})}
}

\tikzset{
  mycallout/.style={
	 fill=gray!30, opacity=.5, overlay, align=center,
	 cloud callout, cloud puffs=10, aspect=1.9, cloud ignores aspect, cloud puff arc=100
  }
}

\newcommandx{\ggvisually}[8][1=5pt,2=15pt,3=5pt,4=5pt,5=1.0cm,6=\scriptsize,7={},8={},usedefault=@]{
  \def\dist{\hspace{#5}}
  $\begin{array}{c@{\dist}c@{\dist}c@{\dist}c@{\dist}c@{\dist}c}
	  \begin{tikzpicture}[node distance=0.9cm and 0.4cm, every node/.style={scale=.7,transform shape}]
		 \node[source] (srcint) {};
		 \node[sink,below=of srcint] (sinkint) {};
		 \node[mycallout, above = .3cm of srcint, xshift=1cm, callout absolute pointer={(srcint.east)}] {source node};
		 \node[mycallout, below = .3cm of sinkint, xshift=1cm, callout absolute pointer={(sinkint.west)}] {sink node};
		 \path[line] (srcint) -- (sinkint);
	  \end{tikzpicture}
	  &
		 \begin{tikzpicture}[node distance=0.9cm and 0.4cm, every node/.style={scale=.7,transform shape}]
			\mkint{}{int}[]
			\mkgraph{int}{int};
		 \end{tikzpicture}
	  &
		 \begin{tikzpicture}[node distance=.9cm and 0.4cm, every node/.style={scale=.7,transform shape}]
			\node[bblock] at (0,0) (g) {$\aG$};
			\node[node, below=of g] (s1) {};
			\node[bblock, below=of s1] (gp) {$\aG'$};
			\path[line,dotted] (g) -- (s1);
			\path[line,dotted] (s1) -- (gp);
		 \end{tikzpicture}
	  &
		 \begin{tikzpicture}[node distance=.4cm and 0.4cm, every node/.style={scale=.7,transform shape}]
			\node[bblock] at (-.7,0) (g) {$\aG$};
			\node[bblock] at (.7,0)  (gp) {$\aG'$};
			\node[node, above=of g] (f) {};
			\node[node, below=of g] (j) {};
			\node[node, above=of gp] (fp) {};
			\node[node, below=of gp] (jp) {};
			\path[line,dotted] (f) -- (g);
			\path[line,dotted] (g) -- (j);
			\path[line,dotted] (fp) -- (gp);
			\path[line,dotted] (gp) -- (jp);
			\mkfork{f,fp}[fork][][#1];
			\mkjoin{j,jp}[join][][#2];
			\mkgraph{fork}{join};
			\node[mycallout, above = .3cm of fork, xshift=-1cm, callout absolute pointer={(fork.west)}] {fork gate};
			\node[mycallout, above = -.9cm of join, xshift=-1cm, callout absolute pointer={(join.west)}] {join gate};
		 \end{tikzpicture}
	  &
		 \begin{tikzpicture}[node distance=.4cm and 0.4cm, every node/.style={scale=.7,transform shape}]
			\node[bblock] at (-.7,0) (g) {$\aG$};
			\node[bblock] at (.7,0)  (gp) {$\aG'$};
			\node[node, above=of g] (f) {};
			\node[node, below=of g] (j) {};
			\node[node, above=of gp] (fp) {};
			\node[node, below=of gp] (jp) {};
			\path[line,dotted] (f) -- (g);
			\path[line,dotted] (g) -- (j);
			\path[line,dotted] (fp) -- (gp);
			\path[line,dotted] (gp) -- (jp);
			\mkbranch{f,fp}[fork][][#3];
			\mkmerge{j,jp}[join][][#4];
         \mkgraph{fork}{join};
         \node[mycallout, above = .3cm of fork, xshift=-1cm, callout absolute pointer={(fork.west)}] {branch gate};
         \node[mycallout, above = -.9cm of join, xshift=-1cm, callout absolute pointer={(join.west)}] {merge gate};
       \end{tikzpicture}
     \ifempty{#7}{}{
     &
		 \begin{tikzpicture}[node distance=0.4cm and 0.4cm, every node/.style={scale=.7,transform shape}]
        \node[bblock] (g) {$\aG$};
        \node[node, above=.5cm of g] (f) {};
        \node[node, below=.5cm of g] (j) {};
        \path[line,dotted] (f) -- (g);
        \path[line,dotted] (g) -- (j);
        \mkloop[.4][1]{f}{j};
        \mkgraph[.3cm]{entryf}{exitj};
        \node[mycallout, above = .2cm of entryf, xshift=1.3cm, callout absolute pointer={(entryf.east)}] {loop entry};
        \node[mycallout, above = -.7cm of exitj, xshift=1.3cm, callout absolute pointer={(exitj.west)}] {loop exit};
      \end{tikzpicture}
     }
     \ifempty{#8}{}{
	  \\
     \text{#6 empty}
     &
     \text{#6 interaction}
     &
     \text{#6 sequential}
     &
     \text{#6 parallel}
     &
     \text{#6 branch}
		 &\text{#6 iteration}
   }
   \end{array}$
}

\newcommandx{\newggvisually}[5][1=5pt,2=15pt,3=\scriptsize,4={},5={},usedefault=@]{
  \def\w{1cm}
  \begin{minipage}[c]{\w}
	 \ifempty{#5}{}{\text{#3 empty}\\[#1]}
	 \begin{tikzpicture}[node distance=0.9cm and 0.4cm, every node/.style={scale=.7,transform shape}]
		\node[source] (srcint) {};
		\node[sink,below=of srcint] (sinkint) {};
		\node[mycallout, above = .3cm of srcint, xshift=1cm, callout absolute pointer={(srcint.east)}] {source node};
		\node[mycallout, below = .3cm of sinkint, xshift=1cm, callout absolute pointer={(sinkint.west)}] {sink node};
		\path[line] (srcint) -- (sinkint);
	 \end{tikzpicture}
  \end{minipage}
  \hfill
	  \begin{minipage}[c]{\w}
     \ifempty{#5}{}{\text{#3 interaction}\\[#1]}
		 \begin{tikzpicture}[node distance=0.9cm and 0.4cm, every node/.style={scale=.7,transform shape}]
			\mkint{}{int}[]
			\mkgraph{int}{int};
		 \end{tikzpicture}
	  \end{minipage}
	 \hfill
	  \begin{minipage}[c]{.1cm}
     \ifempty{#5}{}{\text{#3 sequential}\\[#1]}
		 \begin{tikzpicture}[node distance=.9cm and 0.4cm, every node/.style={scale=.7,transform shape}]
			\node[bblock] at (0,0) (g) {$\aG$};
			\node[node, below=of g] (s1) {};
			\node[bblock, below=of s1] (gp) {$\aG'$};
			\path[line,dotted] (g) -- (s1);
			\path[line,dotted] (s1) -- (gp);
		 \end{tikzpicture}
	  \end{minipage}
	  \hfill
	  \begin{minipage}[c]{\w}
     \ifempty{#5}{}{\text{#3 parallel}\\[#1]}
		 \begin{tikzpicture}[node distance=.4cm and 0.4cm, every node/.style={scale=.7,transform shape}]
			\node[bblock] at (-.7,0) (g) {$\aG$};
			\node[bblock] at (.7,0)  (gp) {$\aG'$};
			\node[node, above=of g] (f) {};
			\node[node, below=of g] (j) {};
			\node[node, above=of gp] (fp) {};
			\node[node, below=of gp] (jp) {};
			\path[line,dotted] (f) -- (g);
			\path[line,dotted] (g) -- (j);
			\path[line,dotted] (fp) -- (gp);
			\path[line,dotted] (gp) -- (jp);
			\mkfork{f,fp}[fork][][#1];
			\mkjoin{j,jp}[join][][#2];
			\mkgraph{fork}{join};
			\node[mycallout, above = .2cm of fork, xshift=-1cm, callout absolute pointer={(fork.west)}] {#3 fork gate};
			\node[mycallout, below = .2cm of join, xshift=-1cm, callout absolute pointer={(join.west)}] {#3 join gate};
		 \end{tikzpicture}
	  \end{minipage}
	  \hfill
	  \begin{minipage}[c]{\w}
     \ifempty{#5}{}{\text{#3 branch}\\[#1]}
		 \begin{tikzpicture}[node distance=.4cm and 0.4cm, every node/.style={scale=.7,transform shape}]
			\node[bblock] at (-.7,0) (g) {$\aG$};
			\node[bblock] at (.7,0)  (gp) {$\aG'$};
			\node[node, above=of g] (f) {};
			\node[node, below=of g] (j) {};
			\node[node, above=of gp] (fp) {};
			\node[node, below=of gp] (jp) {};
			\path[line,dotted] (f) -- (g);
			\path[line,dotted] (g) -- (j);
			\path[line,dotted] (fp) -- (gp);
			\path[line,dotted] (gp) -- (jp);
			\mkbranch{f,fp}[fork][][#1];
			\mkmerge{j,jp}[join][][#2];
         \mkgraph{fork}{join};
         \node[mycallout, above = .1cm of fork, xshift=-1cm, callout absolute pointer={(fork.west)}] {branch gate};
         \node[mycallout, below = .1cm of join, xshift=-1cm, callout absolute pointer={(join.west)}] {merge gate};
       \end{tikzpicture}
	  \end{minipage}
     \ifempty{#4}{}{
     \hfill
	  \begin{minipage}[c]{\w}
     \ifempty{#5}{}{\text{#3 iteration}\\[#1]}
		 \begin{tikzpicture}[node distance=0.4cm and 0.4cm, every node/.style={scale=.7,transform shape}]
        \node[bblock] (g) {$\aG$};
        \node[node, above=.5cm of g] (f) {};
        \node[node, below=.5cm of g] (j) {};
        \path[line,dotted] (f) -- (g);
        \path[line,dotted] (g) -- (j);
        \mkloop[.4][1]{f}{j};
        \mkgraph[.3cm]{entryf}{exitj};
        \node[mycallout, above = .2cm of entryf, xshift=1.3cm, callout absolute pointer={(entryf.east)}] {loop entry};
        \node[mycallout, above = -.7cm of exitj, xshift=1.3cm, callout absolute pointer={(exitj.east)}] {loop exit};
      \end{tikzpicture}
	  \end{minipage}
     }
}

  \newcommandx{\wwwcquote}[1][1=quo:w3c,usedefault=@]{
	 \ifempty{#1}{}{\begin{quote}\label{#1}}
		\lq\lq Using the Web Services Choreography specification, a
		\textcolor{orange}{contract} containing a global definition of the
		common \textcolor{orange}{ordering} conditions and constraints
		under which \textcolor{orange}{messages} are exchanged, is
		produced that describes, from a \textcolor{orange}{global
		  viewpoint} [...]  observable behaviour of all the parties
		involved.
		\textcolor{OliveGreen}{Each party} can then use the global definition to
		\textcolor{OliveGreen}{build and test solutions that conform to it}.
		The global specification is in turn \textcolor{OliveGreen}{realised by combination of} the
		resulting \textcolor{OliveGreen}{local systems} [...]\rq\rq
		\ifempty{#1}{}{\end{quote}}
  }

\newcommand{\varset}{\mathtt{X}}
\newcommandx{\constset}[1][1=C,usedefault=@]{\mathsf{#1}}
\newcommandx{\termalg}[2][1=\Sigma,2=\varset,usedefault=@]{\mathsf{Term}_{#1,#2}}
\newcommandx{\formulas}[2][1=\Sigma,2=\varset,usedefault=@]{\mathsf{Form}_{#1,#2}}
\newcommandx{\modelsclass}[2][1=\Sigma,2=\varset,usedefault=@]{\mathsf{Mod}_{#1,#2}}

\newcommand{\QL}{\mathcal{QL}}

\newcommand{\DLTL}{\mathit{DLTL}}
\newcommand{\gqosprop}{\Phi}

\mkfun{\aggfn}{agg}{}
\mkfun{\qos}{qos}{}
\mkfun{\splitfn}{split}{}

\def \({\left (}
\def \){\right )}
\def \[{\left [}
\def \]{\right ]}
\def \[[{\left \llbracket}
\def \]]{\right \rrbracket}
\def \<{\left\langle}
\def \>{\right\rangle}

\newcommandx{\psiinst}[3][1=\psi,2=a,3=q,usedefault=@]{{#1}_{\ptp[#2]}^{{#3}}}
\let\vec\mathfrak

\newcommand{\thetool}{\toolid{MoCheQoS}}
\newcommand{\qluntil}[1][\aG]{\;\mathbf{\mathsf{U}}^{#1}\;}
\makeatletter
\newcommand{\dolist}[2]{%
  \def\nextitem{\def\nextitem{#1}}%
  \@for \el:=#2\do{\nextitem\textbf{\el}}%
}
\makeatother
\newcommand{\qosattr}[1]{\ensuremath{\mathsf{#1}}}

\def\captionsep{-15pt}
\title{
  \thetool: Automated Analysis of Quality of Service Properties of Communicating Systems
   \thanks{Part of the material in this paper has been published at the $26^{th.}$ International Symposium on Formal Methods -- {FM 2024} where the companion artifact was awarded the \emph{Available} and \emph{Reusable} badges.}
}
\titlerunning{\thetool: Analysis of QoS Properties of Communicating Systems}

\author{Carlos G. Lopez Pombo \inst{1}\thanks{On leave from Instituto de Ciencias de la
   	 computación CONICET--UBA and Departamento de Computación,
   	 Facultad de Ciencias Exactas y Naturales, Universidad de Buenos
  	 Aires.
  }, 
  Agust\'in E. Martinez Su\~n\'e\inst{2}, 
  Emilio Tuosto\inst{3}
}

\institute{%
  Centro Interdisciplinario de Telecomunicaciones, Electrónica,
  Computación y Ciencia Aplicada, Universidad Nacional de
  Río Negro - Sede Andina and CONICET
  \and
  CONICET--UBA. Instituto de Investigaci\'on en Ciencias de la Computaci\'on 
  \and 
  Gran Sasso Science Institute
}

\authorrunning{%
  Carlos G. Lopez Pombo, Agust\'in E. Martinez Su\~n\'e, Emilio Tuosto
}
\index{Lopez Pombo, Carlos G.}
\index{Martinez Su\~n\'e, Agust\'in E.}
\index{Tuosto, Emilio}

\begin{document}

\maketitle

\begin{abstract}
  We present \thetool, a bounded \textsf{\textcolor{teal}{mo}}del
  \textsf{\textcolor{teal}{che}}cker to analyse (\toolid{QoS}) properties of
  message-passing systems.
  Building on the dynamic temporal logic, the choreographic model, and
  the bounded model checking algorithm defined in our ICTAC 2023
  paper, \thetool enables the static analysis of QoS properties of
  systems built out from the composition of services.
  We consider QoS properties on measurable application-level
  attributes as well as resource consumption metrics, for example those
  relating monetary cost to memory usage.
  The implementation of the tool is accompanied by an experimental
  evaluation.
  More precisely, we present two case studies meant to evaluate the
  applicability of \thetool; the first is based on the AWS cloud while
  the second analyses a communicating system automatically extracted
  from code.
  Additionally, we consider synthetically generated experiments to
  assess the scalability of \thetool.
  These experiments showed that our model can faithfully capture and
  effectively analyse QoS properties in industrial-strength scenarios.
\end{abstract}

\section{Introduction}\label{sec:intro}
%!TEX root = ./main.tex
Monolithic applications are steadily giving way to distributed
cooperating components implemented as services.
This transition was accelerated by the software-as-a-service motto
triggered in the 21$^\text{st}$ century by the service-oriented
computing (SOC) paradigm, later evolved in e.g., cloud, fog, or edge
computing.
These paradigms envisage software systems as applications running over
globally available computational and networking infrastructures to
procure services that can be composed on the fly so that, collectively,
they can fulfil given goals \cite{fiadeiro:fac-23_4}.

Key to this trend are \emph{Service Level Agreements} (SLAs) that
express the terms and conditions under which services interact.
An essential element covered by SLAs are the quantitative constraints
specifying non-functional behaviour of services.
For example, the current SLA and pricing scheme for the AWS Lambda
service~\cite{aws:sla,aws:price} declare constraints on quantifiable
attributes. %  prior to the service use.
To the best of our knowledge, the standard practice is to
informally specify the SLA of each service
provided and then use run-time verification (like monitoring) to check
quantitative non-functional properties.
This approach makes it difficult to check system-level properties
because SLAs (besides being informal) do not specify conditions on the
composition of services.

Since their introduction in~\cite{w3c:wsdl20}, choreographies stood
out for a neat separation of concerns: choreographic models abstract
away local computations focusing on the communications among
participants; therefore, they are spot on for services
since they reconcile the \squo{atomistic} view at the services'
interactions level with the \squo{holistic} view at the system level.
Indeed, choreographies require to specify a high-level description of
interactions (the \emph{global view}) and relate it to a description
of services' behaviour (the \emph{local view}).
%
% %
% Choreographies are gaining momentum in industry
% (e.g.~\cite{bpmn,bon18,fmmt20,DBLP:journals/software/AutiliIT15})
% which, increasingly, conceives applications as \emph{communicating}
% services.
% 
These are distinctive features of the choreographic framework presented
in~\cite{lopezpombo:ictac23} to provide reasoning capabilities about the
QoS of communicating systems, starting from the QoS of the underlying services.
The basic idea in~\cite{lopezpombo:ictac23} is:
\begin{inparaenum}[(i)]
\item to specify constraints on quality attributes of local states of
  services and
\item to verify through a bounded model-checking algorithm system-level QoS
  properties expressed in $\QL$, a specific dynamic logic where
  temporal modalities are indexed with \emph{global choreographies}
 (g-choreographies~\cite{tuosto:jlamp-95}, a formal model of global
 views of choreographies).
\end{inparaenum}
A simple example can illustrate this.
Let \p\ be a service that converts files to various formats and
invokes a storage service \q\ to $\msg[s]$ave the results of requests;
both \p\ and \q\ charge customers depending on the size of stored data
(as done e.g. by Amazon's DynamoDB service).
The request of \p\ to \q\ can be abstracted away with two finite-state
machines whose states are decorated with constraints on
the two quality attributes: monetary cost ($\qosattr{c}$) and data
size ($\qosattr{s}$) as follows: 
\begin{align*}\label{eq:idea}
  \tikzset{
  intronode/.style={
  font=\scriptsize,
  node distance =1.5cm
  },
  introlabel/.style={
  scale=.75,color=blue,fill=yellow!10
  }
  }
  \begin{tikzpicture}[every node/.style=intronode, every label/.style=introlabel]
	 \node[label={left:$\left\{
		\begin{array}[c]{c}
		  \qosattr{c} \leq 5,\\ \qosattr{s} = 0
		\end{array}
		\right\}$}] (aq0) {$\aQzero$};
    \node[right = of aq0, label={right:$\left\{
		  \begin{array}[c]{c}
			 5 \leq \qosattr{c} \leq 10,
			 \\
			 \qosattr{s} < 3 
		\end{array}
	 \right\}$}] (aq1){$q_1$};
	 \path[->,draw] (aq0) edge node[above] {$\aout[@][@][s]$} (aq1);
  \end{tikzpicture}
  \qand
  \begin{tikzpicture}[every node/.style=intronode, every label/.style=introlabel]
    \node[label={left: $\left\{
		  \begin{array}[c]{c}
			 \qosattr{c}=0,
			 \\
			 \qosattr{s} = 0
		\end{array}
		\right\}$}] (bq0){$\aQzero'$};
    \node[right = of bq0,label={
		right:$\left\{\begin{array}[c]{c} 10 \leq \qosattr{s} \leq 50,\\ \qosattr{c} = 0.01 \cdot \qosattr{s}
		\end{array}
	 \right\}$
  }] (bq1){$q_1'$};
	 \path[->,draw] (bq0) edge node[above] {$\ain[@][@][s]$} (bq1);
  \end{tikzpicture}
\end{align*}
Intuitively, the formulae associated to states constraint the quality
attributes upon the local computation executed in the states.
For instance, both services store no data in their initial state;
computation in \p\ may cost up to five units before the request to \q,
which has no cost ($\qosattr{c} = 0$) in $q_0'$ since it is just
waiting to execute the input.
If, as we assume, communication is asynchronous,
then the composition of \p\ and \q\ yields a run like
$\pi : \aConf_0 \xrightarrow{\aout[@][@][s]} \aConf_1
\xrightarrow{\ain[@][@][s]} \aConf_2$ where first \p\ sends the
message and then \q\ received it.
Then, the system-level QoS of the composition of \p\ and \q\ would be the result of
\emph{aggregating} the constraints on $\qosattr{c}$ and $\qosattr{s}$
along the run $\pi$.

% \medskip

\noindent \textbf{Structure \& Contributions}
  A main contribution of this paper is a tool to support the static
  analysis technique of QoS properties of message-passing systems.
  More precisely, we implement the bounded model-checking algorithm
  introduced in~\cite{lopezpombo:ictac23} (and summarised in
  \cref{sec:pre}) in a tool called \thetool (after
  \textsf{\textcolor{teal}{Mo}}del-\textsf{\textcolor{teal}{Che}}cker
  for \textsf{\textcolor{teal}{QoS}} properties).%
By combining the SMT solver Z3~\cite{demoura:tacas08} and the
choreographic development toolchain
\chorgram~\cite{cgtCOORD20,chorgram,lty17} (as discussed in
\cref{sec:tool}), \thetool can model-check QoS properties expressed in
$\QL$, the dynamic temporal logic of~\cite{lopezpombo:ictac23}.
\thetool is publicly available at \cite{lopezpombo24}.

A key feature of our approach is that the analysis of QoS properties
of systems builds on the QoS constraints specified on the components
of the system; as seen in the example above, \thetool features the capability of aggregating QoS
constraints along the computation of systems.

Another contribution is the empirical evaluation of our approach (\cref{sec:eval}),
which is done through:
\begin{inparaenum}[($a$)]
\item a case study borrowed from the AWS Cloud~\cite{aws:cloud},
\item a case study borrowed from the literature~\cite{imai:tacas22}
  where communication protocols are automatically extracted from code, and
\item synthetic examples designed to evaluate the scalability of
  \thetool.
\end{inparaenum}

After discussing related work (\cref{sec:rw}), we draw conclusions and
suggest future work (\cref{sec:conclu}).
Auxiliary material is relegated to appendixes.

%%% Local Variables:
%%% mode: latex
%%% TeX-master: "main"
%%% End:

\section{Preliminaries}\label{sec:pre}
	%!TEX root = ./main.tex
We fix a set $\PSet$ of \emph{participants} (identifying interacting
services) and a set $\msgset$ of
(types of) \emph{messages} such that $\PSet \cap \msgset = \emptyset$.
The communication model of \thetool hinges on \emph{QoS-extended
  communicating finite-state machines}~\cite{lopezpombo:ictac23}
(qCFSMs for short).
A CFSM~\cite{brand:jacm-30_2} is a finite state automaton whose
transitions are labelled by output or input actions.
An output action $\aout$ (resp. input action $\ain$) specifies the
output (resp. input) of a message $\msg$ from \p\ to \q\ (resp. received by
\q\ from \p).
A qCFSM is a CFSM where \emph{QoS specifications}, that is first-order
formulae predicating over QoS attributes, decorate
states.\footnote{Unlike CFSMs, qCFSMs feature \emph{accepting} states,
represented here as double circles.}

\begin{example}\label{ex:simple-cfsm-qos}
  \def\specA{\Gamma_\mathrm{A}}
  \def\specB{\Gamma_\mathrm{B}}
  Let
  $\specA=\{5 \leq \qosattr{mem} \leq 10, \qosattr{cost} = 0.2 \cdot
  \qosattr{mem}\}$ and $\specB=\{\qosattr{mem} = 0, \qosattr{cost} = 1\}$
  be two QoS specifications.
  In the system made of the qCFSMs
  \\
  \begin{tikzpicture}[node distance=1cm and 1.2cm, initial where=left, transform shape, every label/.style={color=blue,fill=yellow!10},scale=.9, transform shape]
	 \node[cnode, initial,  initial text={$M_\text{\p}$:}] (A0) {0};
	 \foreach \s/\l/\t/\i in {0/\specA/1/1,1/{}/2/2}{
		\node (A\t) [cnode, right = of A\s, label = $\l$] {\i};
	 }
	 \node (A3) [cnode, right = of A2, double] {3};
	 \foreach \s/\l/\t in {0/\aout[a][b][x]/1,2/\aout[a][b][z2]/3}{
		\path[line,sloped] (A\s) edge node[below]{$\l$} (A\t);
	 }
	 \path[line] (A1) edge[bend left=-35] node[below]{$\ain[b][a][y]$} (A2);
	 \path[line] (A2) edge[bend left=-35] node[above]{$\aout[a][b][z1]$} (A1);
	 \node[cnode, right = 2cm of A3, initial, initial text={$M_\text{\q}$:}] (B0) {0};
	 \foreach \s/\l/\t/\i in {0/{}/1/1,1/{}/2/2}{
		\node (B\t) [cnode, right = of B\s] {\i};
	 }
	 \node (B3) [cnode, right = of B2, label = $\specB$, double] {3};
	 \foreach \s/\l/\t in {0/\ain[a][b][x]/1,2/\ain[a][b][z2]/3}{
		\path[line,sloped] (B\s) edge node[below]{$\l$} (B\t);
	 }
	 \path[line] (B1) edge[bend left=-35] node[below]{$\aout[b][a][y]$} (B2);
	 \path[line] (B2) edge[bend left=-35] node[above]{$\ain[a][b][z1]$} (B1);
  \end{tikzpicture}
  \\
  participant \p\ first sends message $\msg[x]$ to \q, then \q\ and
  \p\ exchange messages $\msg[y]$ and $\msg[z1]$ an unbounded number
  of times, and finally \p\ sends message $\msg[z2]$ to \q.
  \finex
\end{example}

A \emph{QoS-extended communicating system} (qCS for short) is a map
assigning a qCFSM to participants in $\PSet$; for instance, the map
$\aCS$ where $\aCS(\p) = \aCM_{\p}$ and $\aCS(\q) = \aCM_{\q}$ are the
qCFSM of \cref{ex:simple-cfsm-qos} is a communicating system.
Since QoS specifications do not affect communications, the semantics
of qCSs is as the one of communicating systems.
Let us recall how CFSMs interact.

Communicating systems are asynchronous: the execution of an output
action $\aout$ allows the sender \p\ to continue even if the receiver
\q\ is not ready to receive; message $\msg$ is appended in an infinite
FIFO buffer, the \emph{channel} $\achan$, from where \q\ can consume
$\msg$.
Formally, given a communicating system $\aCS$ on $\ptpset$, we define
a labelled transition system (LTS) whose transitions relate
\emph{configurations} and communication actions.
A configuration is a pair $\csconf q \abuffer$ where $\vec q$ and
$\vec \abuffer$ respectively maps each participant \p\ to a state of
$\aCS(\p)$ and each channel to a sequence of messages; state
$\vec q(\p)$ keeps track of the state of machine $\aCS(\p)$ and buffer
$\abuffer(\achan)$ yields the messages sent from \p\ to \q\ and not
yet consumed.
Let $\aConf_0$ denote the \emph{initial} configuration where, for all
$\p \in \ptpset$, $\vec q(\p)$ is the initial state of $\aCS(\p)$ and
$\abuffer(\achan)$ is the empty sequence for all channels $\achan$.

A configuration $\csconf q \abuffer$ \emph{reaches} another
configuration $\csconf{q'} {\abuffer'}$ \emph{with a transition $\al$}
if there is a message $\msg \in \msgset$ such that either (1) or (2)
below holds:
%
% A configuration $\aConf'= \csconf {q'} {\abuffer'} $ is {\em
%   reachable} from another configuration
% $\aConf = \csconf {q} {\abuffer}$ by \emph{firing a transition $\al$},
% written $\aConf \TRANSS{\al} \aConf'$, if there is a message
% $\msg \in \msgset$ such that either (1) or (2) below holds:
\begin{center}
  \begin{tabular}{l@{\hfill}r}
    \begin{minipage}{.47\linewidth}\small
      1.  $\al = \aout[@][@][@]$ with
		$\vec q(\p) \trans[\p]{\al} {q'}$ and\vspace{-7pt}
        \begin{itemize}
        \item[a.] $\vec q' = \vec q[\p \mapsto q']$
        \item[b.] $\abuffer' = \abuffer[\achan \mapsto \abuffer(\achan).\msg]$
        \end{itemize}
    \end{minipage}
    &
    \begin{minipage}{.47\linewidth}\small
      2.
        $\al = \ain[@][@][@]$ with $\vec q(\q) \trans[\q]{\al} {q'}$
        and \vspace{-7pt}
        \begin{itemize}
        \item[a.] $\vec q' = \vec q[\q \mapsto q']$ and
        \item[b.] $\abuffer = \abuffer'[\achan \mapsto \msg.\abuffer'(\achan)]$
        \end{itemize}
    \end{minipage}
  \end{tabular}
\end{center}
in (1) $\msg$ is sent on $\achan$ while in (2) it is received.
Machines and buffers not involved in the transition are left unchanged.
We write
$\aConf \TRANSS\al \aConf'$ when $\aConf$ reaches $\aConf'$.

Let $\aCS$ be a communicating system,
a sequence $\pi = (\aConf_i,\al_i,\aConf_{i+1})_{i \in I}$ where
$I$ is an initial segment of natural numbers (i.e., $i-1 \in I$ for
each $0 < i \in I$) is a run of $\aCS$ if
$\aConf_i \TRANSS{\al_i} \aConf_{i+1}$ is a transition of $\aCS$ for
all $ i \in I$.
The set of runs of $\aCS$ is denoted as $\Delta^\infty_{\aCS}$
and the set of runs of length $k$ is denoted as $\Delta^k_{\aCS}$.
Note that $\Delta^\infty_{\aCS}$ may contain runs of infinite length, 
the set of finite runs of $\aCS$ is the union of all $\Delta^k_{\aCS}$ 
and will be denoted as $\Delta_{\aCS}$.
Given a run $\pi$,
%$ = (\aConf_i,\al_i,\aConf_{i+1})_{i \in I}$
 we define
$\rlang[\pi]$ to be the sequence of labels $(\al_i)_{i \in I}$.
The \emph{language} of $\aCS$ is the set
$\rlang[\aCS] = \{\rlang[\pi] \sst \pi \in
\Delta^\infty_{\aCS}\}$.
Finally, $\mathit{prf}: \Delta^\infty_{\aCS} \to 2^{\Delta_{\aCS}}$
maps each run $\pi \in \Delta^\infty_{\aCS}$ to its set of finite
prefixes.
As usual, for all $\pi \in \Delta^\infty_{\aCS}$, the empty prefix
$\epsilon$ belongs to $\mathit{prf} (\pi)$.
For convenience, we will occasionally write
$ \aConf_0 \TRANSS{\al_0} \aConf_1\ \ldots\
\aConf_n \TRANSS{\al_n} \aConf_{n+1}$ for finite sequences.

The logic $\QL$ is introduced in~\cite{lopezpombo:ictac23} to express
system-level QoS properties.
Akin $\DLTL$~\cite{henriksen:apal-96_1_3}, $\QL$ is basically a linear
temporal logic where atomic formulae, ranged over by $\psi$,
constrain quantitative attributes, and the \squo{until} modality
is restricted to specific runs.
The syntax of $\QL$ is given by the grammar:
\begin{align*}
  \gqosprop & \bnfdef
				  \top \bnfmid
				  \psi \bnfmid
				  \neg \gqosprop \bnfmid
				  \gqosprop \lor \gqosprop \bnfmid
				  \gqosprop\ \qluntil \gqosprop
\end{align*}
where $\top$ is the truth value 'true', $\neg$ and $\lor$ are the
usual connectives for logical negation and disjunction, and the index
$\aG$ of the \squo{until} modality is a \emph{global choreography}
(g-choreographies for short)~\cite{tuosto:jlamp-95} meant to restrict
the set of runs to be considered for the satisfiability of formulae\footnote{
  Logical connectives $\land$ and $\implies$ are defined as usual
  while possibility $\possib \aG \gqosprop$ and necessity
  $[\aG] \gqosprop$ are defined as $\top \qluntil \gqosprop$
  and $\neg \possib \aG \neg \gqosprop$ respectively.
}.
G-choreographies can be thought of as regular expressions on the alphabet
$\{\gbrkkw\} \cup \{ \gint[] \sst \p,\q \in \PSet, \msg \in \msgset\}$
where, $\gbrkkw$ is used to stop
iterations and $\gint[]$ represents an interaction where \p\ and \q\
exchange message $\msg$.
We let $\gcho[][\_][\_]$, $\grec[][\_]$, and $\gseq[][\_][\_]$
respectively denote choice operator, Kleene star, and sequential
composition (with $\gseq[][\_][\_]$ taking precedence over
$\gcho[][\_][\_]$).
\begin{example}\label{ex:gc-pop}
  The g-choreography
  $\aG_\mathsf{sys} = \gint[][a][x][b] \gseqop \gint[][b][y][a]
  \gseqop {\aG_{\mathsf{exch}}}^{\grecop} \gseqop \gint[][a][z2][b]$
  corresponds to the qCS in \cref{ex:simple-cfsm-qos} with
  $\aG_{\mathsf{exch}} = \gint[][a][z1][b] \gseqop \gint[][b][y][a]$
  specifying the exchange of messages $\msg[z1]$ and $\msg[y]$ between
  \p\ and \q.
  \finex
\end{example}
A g-choreography $\aG$ induces a causality relation on the underlying
communication whereby the output of an interaction precedes the
corresponding input and, for the sequential composition $\aG ; \aG'$
the actions in $\aG$ precede those in $\aG'$ when executed by a same
participant.
The \emph{language} $\rlang[\aG]$ of a g-choreography $\aG$ consists
of all possible sequences of communication actions compatible with the
causal relation induced by $\aG$ (note that $\rlang[\aG]$ is
prefix-closed).
We write $\hat{\rlang}[\aG]$
for the set of sequences in $\rlang[\aG]$ that are not proper prefixes
of any other sequence in $\rlang[\aG]$.
The technical definition of $\rlang[\aG]$, immaterial here,
% requires a formal semantics for the g-choreography, we 
% present in \cref{sec:sem} a definition that rely on the pomset semantics of g-choreographies that
can be found in~\cite{tuosto:jlamp-95}.\footnote{%
  As an example, consider $\aCS$ from and \cref{ex:simple-cfsm-qos}
  and $\aG_\mathsf{sys}$ from \cref{ex:gc-pop}; we have
  $\rlang[\aG_\mathsf{sys}] \subset \rlang[\aCS]$ since
  $\aG_\mathsf{sys}$ requires that $\msg[z1]$ is sent at least one
  time, while $\aCS$ allows executions where $\msg[z1]$ is not
  exchanged.
}
% ~\cite{pratt:ijpp-15_1,tuosto:jlamp-95}.
% 
% We will adapt CFSM~\cite{brand:jacm-30_2} to model the QoS-aware \emph{local
% view} of a system.
% 
The next example builds on \cref{ex:simple-cfsm-qos} to illustrate how to express a QoS
property in $\QL$.
\begin{example}[QoS properties]\label{ex:qos-prop}
  The runs of the system where where \p\ and \q\
  exchange message $\msg[z1]$ and $\msg[y]$ three times
  can be specified by the g-choreography
	 $\aG_3 = \gint[][a][x][b]  \gseqop \gint[][b][y][a] \gseqop \aG_{\mathsf{exch}} \gseqop \aG_{\mathsf{exch}} \gseqop \aG_{\mathsf{exch}}$
  where $\aG_{\mathsf{exch}}$ is defined in \cref{ex:gc-pop}.
  Then the $\QL$ formula
  $\gqosprop \equiv [\aG_3](\qosattr{cost} > 0 ) \implies
  [\gseq[][\aG_3][{\grec[][\aG_{\mathsf{exch}}]}]]\big(\qosattr{cost} \leq \qosattr{mem} \cdot 10
  \big)$ holds either if the first three exchanges do not have positive 
  cost or if the
  cost of every subsequent exchange falls within the
  specified bounds.%
  \finex
\end{example}

The models of $\QL$ are defined in terms of runs of a QoS-extended
communicating systems and an \emph{aggregation
  function}~\cite{lopezpombo:ictac23} that formalises the conditions
for a QoS property to hold in a run.
The aggregation function, denoted below as $\aggfn_{\aCS}$ depends on
application-dependent binary \emph{aggregation operators} that
define how QoS attributes accumulate along a run.
Hereafter, we assume that each QoS attribute has an associated
aggregation operator.

A configuration is \emph{accepting} if all participants are in a
accepting state; a \emph{completion} of a run $\pi$ of a system $\aCS$
is a sequence $\pi'$ ending in an accepting configuration such that
$\pi\pi' \in \Delta^\infty_{\aCS}$.
  A QoS property $\gqosprop$ is \emph{satisfiable} in $\aCS$ if there
  exists a run $\pi \in \Delta^\infty_{\aCS}$ with an accepting
  configuration such that
  $\conf{\pi, \epsilon} \models_{\aCS} \gqosprop$ holds, where
  relation $\conf{\_,\_} \models_{\aCS} \_$ is defined as follows:
\begin{align*}
  \begin{array}{rcl}
    \conf{\pi, \pi'} \models_{\aCS} \psi & \mbox{ iff } & \aggfn_{\aCS} (\pi') \vdash_{\mathit{RCF}} \psi \ \ \text{ if } \psi \text{ atomic and } \pi' \in \mathit{prf}(\pi)
	 \\
    \conf{\pi, \pi'} \models_{\aCS} \neg \gqosprop & \mbox{ iff } & \conf{\pi, \pi'} \models_{\aCS} \gqosprop \text{ does not hold}\\ 
    \conf{\pi, \pi'} \models_{\aCS} \gqosprop_1 \lor \gqosprop_2 & \mbox{ iff } & \conf{\pi, \pi'} \models_{\aCS} \gqosprop_1 \mbox{ or } \conf{\pi, \pi'} \models_{\aCS} \gqosprop_2\\ 
    \conf{\pi, \pi'} \models_{\aCS} \gqosprop_1\ \qluntil\ \gqosprop_2 & \mbox{ iff } & \text{there is a completion } \pi'' \text{ of } \pi' \text{ such that }
	 \\
     & &  \rlang[\pi''] \in \hat{\rlang}[\aG],\ \conf{\pi, \pi'\pi''} \models_{\aCS} \gqosprop_2 \text{ and, for all }\\
     & &  \pi''' \in \mathit{prf} (\pi''), \text{if } \pi''' \neq \pi'' \text{ then } \conf{\pi, \pi'\pi'''} \models_{\aCS} \gqosprop_1.
  \end{array}
\end{align*}
To handle atomic formulae, the first clause leverages
\emph{real-closed fields} (RCFs), a decidable formalisation of the
first-order theory of real numbers~\cite[Thm.~37]{tarski:RM-109}, that we use as a
basis for expressing such QoS constraints~\cite{lopezpombo:ictac23};
we interfaced \thetool with the state-of-the-art SMT solver
Z3~\cite{demoura:tacas08} to check the validity of these constraints
(i.e., $\vdash_{\mathit{RCF}}$).
The \squo{until} modality requires $\gqosprop_2$ to hold at some point
along $\pi$, i.e. on a run $\pi'\pi''$ where completion
  $\pi''$ follow run $\pi$, with $\gqosprop_1$ satisfied up to that
point and $\pi''$ compatible with $\aG$. A run $\pi''$ is compatible
with a g-choreography $\aG$ if it belongs to its language
$\hat{\rlang}[\aG]$.

A QoS property $\gqosprop$ is \emph{satisfiable} in $\aCS$ if there
exists a run $\pi \in \Delta^\infty_{\aCS}$ such that
$\conf{\pi, \epsilon} \models_{\aCS} \gqosprop$, and it is
\emph{valid} (denoted as $\models_\aCS \gqosprop$) if, for all runs
$\pi \in \Delta^\infty_{\aCS}$ that contain a final configuration,
$\conf{\pi, \epsilon} \models_{\aCS} \gqosprop$.
The semi-decision algorithm presented in~\cite{lopezpombo:ictac23}
takes as input a QoS property $\gqosprop$, a communicating system
$\aCS$ and a bound $k$, and returns \emph{true} if and only if there
exists a run $\pi$ of length at most $k$ such that $\conf{\pi, \epsilon}
\models_{\aCS} \gqosprop$. The algorithm iterates over all runs of
length at most $k$ and, for each of them, it calls an auxiliary 
procedure that checks whether the run satisfies the QoS property
by recursively following the definition of $\models_{\aCS}$ presented
above. 

%%% Local Variables:
%%% mode: latex
%%% TeX-master: "main"
%%% End:

\section{A bounded Model Checker for QoS}\label{sec:tool}
% !TEX root = ./main.tex

We now present the architecture of \thetool; a detailed presentation
of its command line interface and the relevant file formats is
in the accompanying artefact submission.
A graphical representation of the architecture
\begin{wrapfigure}[17]{r}{6cm}
  \includegraphics[width=0.5\textwidth]{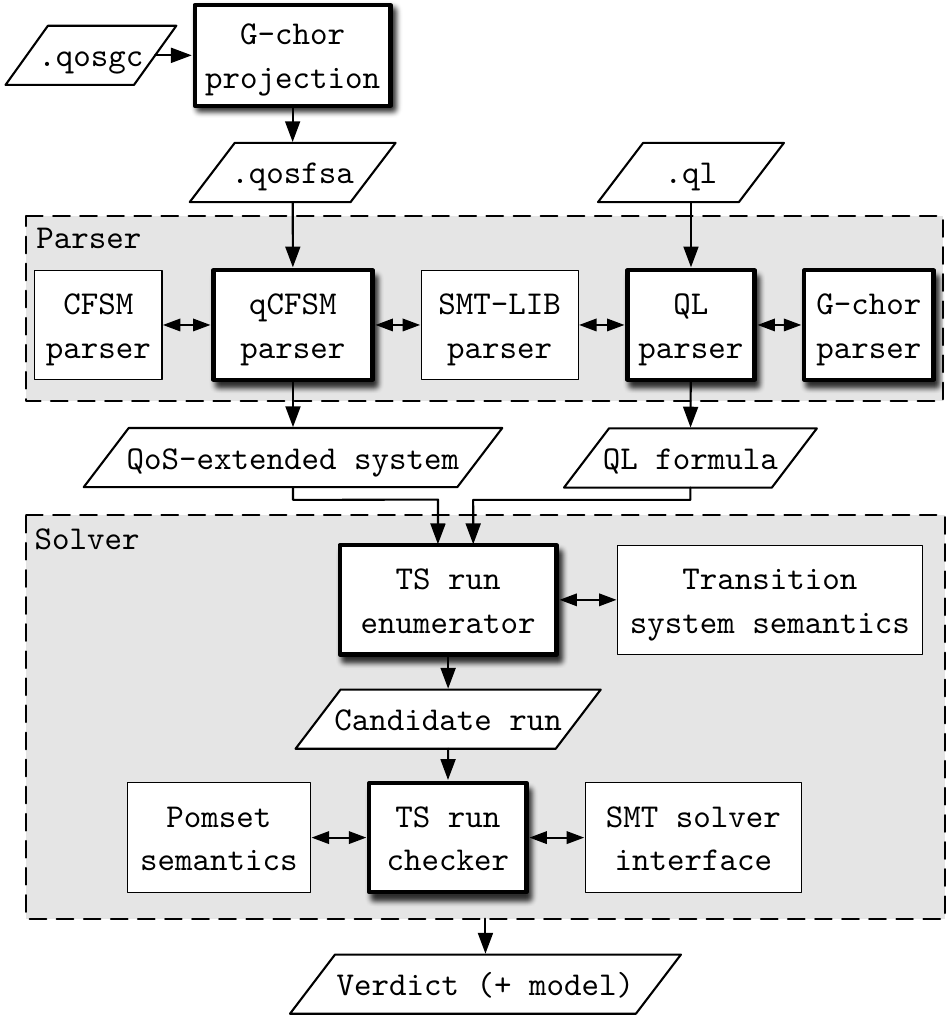}
\end{wrapfigure}
is on the right, where tilted boxes represent files or data objects,
rectangular boxes represent modules or functions, while arrows
represent control and data flows.
Thick shadowed boxes identify the modules developed in this work, while thin
boxes identify \chorgram's modules and other open-source libraries
used by \thetool.
%  (as anticipated, the tool relies on the SMT solver
% Z3~\cite{demoura:tacas08} for the satisfiability of the QoS
% constraints in atomic $\QL$ formulae and on \chorgram for the
% semantics of g-choreographies and CFSMs).
%
% It is worth remarking that we extended the \textsf{G-chor parser}
% of \chorgram\ to support qCFSMs.
%
As \chorgram, \thetool\ is implemented in Haskell.
The two main modules are \texttt{Parser} and \texttt{Solver} (greyed
dashed boxes).
The former transforms the textual representations of \thetool's inputs
into internal Haskell representations used by the latter.
More precisely, \texttt{qCFSM parser} leverages \chorgram's
\texttt{CFSM parser} to process a textual description of the system
from a \texttt{.qosfsa} file.
Likewise, \texttt{QL parser} leverages \texttt{G-chor parser}
(\chorgram's g-choreographies parser) to process a \texttt{.ql} file
containing a simple textual description of the $\QL$ formula to
verify.
Both modules rely on the \texttt{SMT-LIB parser}, Haskell's
\texttt{smt-lib} package.
The format of \texttt{.qosfsa} files is an extension of \chorgram's
\texttt{.fsa} format.
This extension enables the specification of the set of QoS attributes
of interest with ($i$) aggregation operators, ($ii$) QoS constraints
associated to the states of machines, and ($iii$) accepting states of
machines.
Additionally, \thetool\ supports an extension of \chorgram's
g-choreographies to directly specify in a \texttt{.qosgc} file
the QoS-related information on a g-choreography that can be
projected on qCFSMs.
This required to adapt \chorgram's \texttt{G-chor projection} and
\texttt{G-chor parser} modules to support QoS specifications.
 
The \texttt{.ql} format borrows the g-choreography syntax of \chorgram
for the case of the \squo{until} modality.
%
% \eMcomm[(A detailed overview of both \texttt{.qosfsa} and \texttt{.ql} formats can be found in \cref{sec:input-formats}.)]{}
%
The \texttt{Parser} module produces the \texttt{QoS-extended system}
and the \texttt{QL formula} from the input files and passes them to
\texttt{Solver}, our implementation of the bounded model checking
algorithm in~\cite{lopezpombo:ictac23} (cf. \cref{sec:pre}).
More precisely, \texttt{TS run enumerator} invokes \texttt{Transition
  system semantics} (\chorgram's \texttt{TS} module) to incrementally
enumerates the runs of the system (within the bound given in input
parameter \texttt{k}).
Then \texttt{TS run enumerator} invokes \texttt{TS run checker} on
each enumerated run to check if it is a model of the $\QL$ formula.

% The \emph{TS run checker} contains an implementation of the $\QL$
% satisfaction relation that is recursive on the structure of the
% $\QL$ formula.
Properties encompassing sub-formulas of the form $\Phi \qluntil \Phi'$
require \texttt{TS run checker} to invoke \chorgram's
\texttt{PomsetSemantics} module in order to compute the
language\footnote{%
  Iterative sub-terms are replaced by their
  $n$-unfolding, with $0 \leq n \leq \texttt{u}$, where \texttt{u} is
  a an input parameter of \thetool that defaults to parameter \texttt{k}.
  The $n$-unfolding of $\grec[][\aG']$ is
  the sequential composition of $\aG'$ with itself $n$ times.}%
of $\aG$ and to check membership of runs to it. 
Properties involving QoS constraints of an atomic formulae require
\texttt{TS run checker} to invoke the \texttt{SMT solver interface} to
produce and check SMT-LIB queries.
The \texttt{SMT solver interface} is composed of a modified version of
Haskell's \texttt{smt-lib} package to build the SMT-LIB query and of
Haskell's \texttt{SimpleSMT} package to call Z3.
The SMT-LIB query produced for an atomic formula $\psi$ allows to check
whether there exists a counterexample to the entailment
$\aggfn_{\aCS} (\pi') \vdash_{\mathit{RCF}} \psi$.
%
% \eMcomm[move the blue paragraph to artefact?]{}
% \textcolor{blue}{
% This required us to extend %Which is why we modified
% Haskell's \texttt{smt-lib} % by adding support for the
% to support the \texttt{declare-const} command introduced in SMT-LIB 2.5 and for the renaming of symbols in SMT-LIB terms; these features are used by \thetool to declare the QoS attributes and to implement the aggregation function $\aggfn_{\aCS}$, respectively.
% }
% 
% of the aggregated QoS constraints along the run and the QoS constraint in the atomic 
% formula
The SMT-LIB query is then sent to Z3 using Haskell's \texttt{SimpleSMT}
package.
Finally, \texttt{Solver} returns a negative \texttt{Verdict} if the
formula cannot be satisfied within the given bounds or a positive
\texttt{Verdict} with a witnessing \texttt{model} (the run satisfying
the \texttt{QL formula}) otherwise.
% 
% \begin{inparaenum}[(1)]
    % \item the maximum number of unfoldings of the $\grecop$ operator in the g-choreographies, and
    % \item the size of the buffers of the CFSMs.
% \end{inparaenum}
% If the last two parameters are not specified, the tool uses $k$ as the maximum number of unfoldings
% and as the size of the buffers. Since the length of the runs to be considered is always bounded by $k$, 
% these default values are equivalent to not having any bound on the number of unfoldings and on the size
% of the buffers.
% 
  % To avoid recomputing the results of the following operations on the
  % same inputs, \thetool memorises them:
  To optimise computations on same inputs, \thetool memorises the
  results of
\begin{inparaenum}[(a)]
\item the computation of the language of g-choreographies indexing
  \squo{until} operators,
  \label{item:cache-language}
\item the computation of the verdict of atomic entailment $\_ \vdash_{\mathit{RCF}} \_$, and
  % ($\aggfn_{\aCS} (\pi') \vdash_{\mathit{RCF}} \psi$), and
  \label{item:cache-entailment}
\item the membership check of a run to the language of a
  g-choreography.
  \label{item:cache-membership}
\end{inparaenum}
Results of operations \eqref{item:cache-language} and
\eqref{item:cache-membership} are stored using a hash table, while
results of operation \eqref{item:cache-entailment} are stored using a
balanced binary search tree.

%%% Local Variables:
%%% mode: latex
%%% TeX-master: "main"
%%% End:

\section{Evaluation}\label{sec:eval}
Our empirical study aims to evaluate applicability and scalability of
our approach.
Towards applicability, \cref{sec:aws} develops a case study adopting
SLAs from the AWS cloud~\cite{aws:cloud} while \cref{sec:extraction}
borrows a case study from~\cite{imai:tacas22} to show how our approach
can leverage automatic extraction of communicating systems.
Towards scalability, \cref{sec:perf} analyses \thetool\ to measure
its performance.
 
The results presented in the next sections show that our framework can
model SLAs present in industrial-strength scenarios
(\cref{sec:aws}). Notably, \thetool can effectively verify relevant
system-level QoS in such scenarios and produce counterexamples to
refine a property being checked.
Moreover, \cref{sec:extraction} \thetool can be used to effectively
analyse system-level QoS properties of communicating systems
automatically extracted from code.

\subsection{SLA in the Amazon cloud}\label{sec:aws}
%!TEX root = ./main.tex

%\noindent\textbf{SLA in the Amazon cloud}
The case study consists of a three-party version of the POP
protocol~\cite{rfc937} modelled after the OAuth authentication
protocol~\cite{rfc6749}.
More precisely, a client \p[c] securely accesses a remote mailbox
server \p[s] after clearing authentication through a third party
server \p.
This is specified by the g-choreography
\lstinline[language=sgc]{$\aG_{\qosattr{auth}}$ =  $\
  \gint[][c][cred][a]$;$(\gcho[][{\aG_{\qosattr{token}}}][{\gint[][a][error][c]}])$}
where $\aG_{\qosattr{token}}$ models the phase of the OAuth protocol where \p[c]
acquires an authentication token granted if the credentials of the
client \p[c] are valid; the acquired token allows \p[c] to prove its
identity to the POP server \p[s].
This can be modelled as follows:
\begin{lstlisting}[language=sgc, basicstyle=\footnotesize]
  $\aG_{\qosattr{token}}$ = $\ \gint[][a][token][c]$;$\gint[][c][token][s]$;($\gint[][s][fail][c]$ + $\ \gint[][s][ok][c]$;$\aG_{\qosattr{POP}}$)
  $\aG_{\qosattr{POP}}$ = $\ \aG_{\qosattr{quit}}$ + $\ \gint[][c][helo][s]$; $\gint[][s][int][c]$;($\aG_{\qosattr{quit}}$ + $\ \aG_{\qosattr{read}}^{\grecop}$ ; $\aG_{\qosattr{quit}}$)
  $\aG_{\qosattr{quit}}$ = $\ \gseq[][{\gint[][c][quit][s]}][{\gint[][s][bye][c]}]$
  $\aG_{\qosattr{read}}$ = $\ \gint[][c][read][s]$; $\gint[][s][size][c]$;(break + $\ \gint[][c][retr][s]$; $\gint[][s][msg][c]$; $\gint[][c][ack][s]$)
\end{lstlisting}

% \hyphenation{g-chore-og-ra-phy}
We consider a system of qCFSM, one per participant, realising the
g-chore-ography $\aG_{\qosattr{auth}}$
(cf. \cref{appendix:fig:loop-unfoldings-cfsms} in
\cref{appendix:aws}).
The states of the qCFSMs are decorated by QoS specifications derived
by the publicly available SLAs.
Specifically, we use the SLA of the Ory Network identity
infrastructure~\cite{ory.sh} for \p, the one for \p[c] reflects the
SLA of clients of Amazon's Simple Email Service (SES)~\cite{aws:ses},
and the SLA of \p[s] is modelled after the iRedMail service published
in the AWS marketplace~\cite{aws:iredmail}.
We identified the attributes in \cref{attributes:tab}, which are classified as
\emph{cumulative} if attribute values add up some numerical
quantities along the execution, \emph{pricing-scheme} for attributes
specifying parameters for determining the monetary cost of the
service, and \emph{configuration} for attributes fixing the elements
of the computational infrastructure required.
Notice that execution time is measured in seconds, while its
pricing scheme is expressed per hour. Similarly, data transfer
is measured in kilobytes, while its pricing scheme is expressed
per gigabyte. Thus, unit conversions will be performed when computing total
costs.
\begin{table}[t]
\setlength{\belowcaptionskip}{\captionsep}
  \centering
  \begin{tabular}{l|l|l}
	 \multicolumn{1}{c|}{\textbf{cumulative}} & \multicolumn{1}{c|}{\textbf{pricing scheme}} & \multicolumn{1}{c}{\textbf{configuration}}
	 \\\hline
	 num. of emails & price/hour for server software & num. of CPUs 
	 \\
	 data transferred out (Kb) & price/hour for infrastructure usage & amount of memory
	 \\
	 num. of authenticated users & price/Gb for data transfer & network performance
	 \\
	 price for incoming mails & price/user for \p & instance type
	 \\
	 server execution time (s) &&
	 \\
	 total execution time (s) &&
  \end{tabular}
  \caption{SLA attributes for the AWS case study}\label{attributes:tab}
\end{table}
Our approach requires to constrain the quality attributes for each
state of the participant while the constraints specified in the
publicly available SLAs are relative to the whole execution.
% 
% Our qCFMSs provide a more fine-grain view of how the service is executed allowing also a compositional understanding of the costs involved in it.
We overcome this obstacle by identifying the states in the qCFSMs
which are relevant to the constraint and assigning a corresponding
QoS specification to each of these states.
For example, the SLA of Amazon SES specifies that the price paid for
each incoming email is $10^{-4}$ USD; this decorates the state in the
client's qCFSM where mails are received.

The value of the pricing-scheme attributes and the configuration
attributes are determined by the value of the instance type
attribute, which models the type of the \emph{compute}
instance%
\footnote{In AWS jargon, \emph{compute} refers to computational infrastructure, i.e., virtual computers that are rented through services like Amazon Elastic Compute Cloud (EC2).}
decided 
by the user when configuring the services.
%
% In our model, these discrete values are represented with natural
% numbers and the relation between the configuration and the pricing scheme
This relation is rendered in our model with logical implications.
For example, AWS stipulates that 
if the selected instance type is `t4g.nano'%
\footnote{`T4g' is a family of compute instances that are powered by a specific type of processor, while the `nano' suffix indicates the smallest size of the instance.}
(in our model represented by the natural number $1$)
then the hourly rate paid for compute is
$0.0042$ USD; this yields
the following implication to be included in the QoS specification of
the initial state of the server qCFSM together with analogous
implications for other instance types:
\par\nobreak
\vspace*{-1em}
{\fontsize{8}{9}\selectfont
\begin{align*}
   \qosattr{instanceType} = 1 \implies & (\ \qosattr{hrRateCompute} = 0.0042 \land \qosattr{CPUs} = 2 \\ 
  & \land \qosattr{memoryCapacity} = 0.5 \land \qosattr{networkPerformance} \leq 5\ )
\end{align*}
}
% 
% the specification of all the possible configuration can be modeled by a collection of formulae with the shape ``$\left( \bigwedge [ \qosattr{non\_cummulative\_attr} ] = [ \qosattr{natural\_number} ] \right) \implies \left( \bigwedge [ \qosattr{attr} ] = [ \qosattr{value} ] \right)$''.
% 
% For the sake of showing the potential behind our proposal we will take the SLAs available at AWS marketplace and feature a version of it but compatible with the formal language proposed in our work; in addition, we will represent non-cumulative attributes by building a bijection between the discrete choices in the list and natural numbers. Then, the specification of all the possible configuration can be modeled by a collection of formulae with the shape ``$\left( \bigwedge [ \qosattr{non\_cummulative\_attr} ] = [ \qosattr{natural\_number} ] \right) \implies \left( \bigwedge [ \qosattr{attr} ] = [ \qosattr{value} ] \right)$'' therefore, choosing a specific configuration for the service pricing scheme from the available ones boils down to assigning a combination of natural numbers to the corresponding non-cumulative attributes determining the configuration. 
% 
\begin{table}[t]
\setlength{\belowcaptionskip}{\captionsep}
  \centering
  \begin{tabular}{l|c|c}
	 \textbf{} & \textbf{ No. of states } & \textbf{ No. of transitions }
	 \\\hline%\hline
	 Client \p[c]\ & 15 & 17
	 \\
	 Server \p[s]\ & 12 & 14
	 \\
	 Auth. server \p\ & 4 & 3
	 \\
	 LTS of composition\ & 34 & 38
  \end{tabular}
  \caption{Size of the AWS case study model}\label{pop-size:tab}
\end{table}
In summary, the case study is composed of three qCFSMs whose QoS specifications predicate over 14 quality attributes.
The size of the model is reported in \cref{pop-size:tab} and the CFSMs are reported in \cref{appendix:aws}. 
It is worth recalling that CFSMs abstract away from local
computations and focus only on the communication actions. Hence, the
number of states and transitions only reflect the size of the
communication protocol and not necessarily the size of the
implementation. 

Due to space limitations, here we only show the qCFSM for the server \p[s]:
\def\init{\Gamma_\mathrm{init}}
\def\comp{\Gamma_\mathrm{comp}}
\def\data{\Gamma_\mathrm{data}}
  % The qCFSM $\aCM_{\p[c]}$ below specifies the communication behaviour
  % of a POP client \p[c] interacting with a server \p[s]:
\begin{align}
  \aCM_{\p[s]} = &
  \begin{tikzpicture}[node distance=1cm and 1.3cm, scale=.8,transform shape, every label/.style={color=blue,fill=yellow!10,scale=.75}]
	 \node[cnode, initial, initial text = {}, label={$\init$}] (-2) {};
	 \node (-1) [cnode, right = of -2, label={above right:$\comp$}] {};
	 \foreach \s/\l/\t in {-1//0,0/\comp/1,1//2,2/\comp/3,3//4} {
		\node (\t) [cnode, right = of \s, label={above:$\l$}] {};
    }
    \node[cnode, above = of -1] (-3) {};
		\node (5) [cnode, above = of 4, label={above right:$\comp$}] {};
		\node (6) [cnode, left = of 5, label={$\data$}] {};
	 \foreach \s/\l/\t in {-2/\ain[c][s][token]/-1,-1/\aout[s][c][ok]/0,0/\ain[c][s][helo]/1,1/\aout[s][c][int]/2,2/\ain[c][s][read]/3,3/\aout[s][c][size]/4} {
		\path[line,sloped] (\s) edge node[below]{$\l$} (\t);
    }
	 \path[line] (4) edge node[right]{$\ain[c][s][retr]$} (5);
	 \path[line] (5) edge node[above, pos = 0.4]{$\aout[s][c][msg]$} (6);
	 \node[cnode, above = of 0, label={95:$$}] (7) {};
	 \node[cnode, accepting, right = of 7, label={10:$$}] (8) {};
	 \path[line, sloped] (6) edge node[pos=0.35, above]{$\ain[c][s][ack]$} (2);
	 \path[line] (0) edge node[left]{$\ain[c][s][quit]$} (7);
	 \path[line] (2) edge[sloped] node[below,near end]{$\ain[c][s][quit]$} (7);
	 \path[line] (4) edge[sloped] node[pos = 0.6, above = -0.05]{$\ain[c][s][quit]$} (7);
	 \path[line] (7) edge node[above]{$\aout[s][c][bye]$} (8);
	 \path[line] (-1) edge node[left]{$\aout[c][s][fail]$} (-3);
  \end{tikzpicture}
\end{align}
  % The only final state of $\aCM_{\p[c]}$ is the doubly-circled one
  % and the QoS specifications
  % \begin{align*}
  %   & \low = \{t \leq 0.01, c \leq 0.01, m \leq 0.01\},
	%  \\
  %   & \chk = \{ t \leq 5, c = 0.5, m = 0 \}, \\
	% 	&  \{ 1 \leq t \leq 6, c = 0, m \leq 64 \}, \qand
	%  \\
  %   & \db = \{ t \leq 3 \implies (\exists x)(0.5 \leq x \leq 1 \land c = t \cdot x),
  %                                           t > 3 \implies c = 10,
  %                                           m \leq 5 \}
  % \end{align*}
where
$\comp = \{ 0.5 < \qosattr{execTime} < 3, \qosattr{execTimeServer} = \qosattr{execTime} \}$,
modeling states where the server is performing significant computations,
and $\data = \{ 10< \qosattr{dataTransferredOut} < 500 \}$, modeling states where the server has sent data to the client.
The specification $\init$ models the configuration of the AWS instance as stated earlier.
(See full description in the \cref{appendix:aws}).

We now focus on some system-level properties to be checked with \thetool.
In this case study, we are interested in verifying the overall monetary cost of the coordinated execution of the three services. 
By inspecting the pricing scheme in the services SLAs, we can derive the following expression that characterizes the overal cost of an execution of the system in
terms of the aggregated values of the quality attributes:
\par\nobreak
\vspace*{-1em}
{\fontsize{8}{9}\selectfont
\begin{align*}
  \displaystyle
  \qosattr{totalCost} = \ (\qosattr{execTime}/60^2) \cdot \qosattr{hrRateCompute} 
  + (\qosattr{execTimeServer}/60^2) \cdot \qosattr{hrRateServerSoftware}\\
   + \ (\qosattr{dataTransferredOut}/1024^2) \cdot \qosattr{transferGBRate}
   + \qosattr{usersAuth} \cdot \qosattr{ratePerUserAuth}
   + \qosattr{priceEmails}
\end{align*}
}
The values of \qosattr{execTime} and \qosattr{execTimeServer} 
are converted from seconds to hours, and the value of 
\qosattr{dataTransferredOut} from kilobytes to gigabytes.
We can then write the $\QL$ formula
$\Phi_1 = [\aG_\qosattr{init};\aG_\qosattr{msg}]\ \qosattr{totalCost} \leq 1$
where
\begin{align*}
  \aG_{\qosattr{init}} = &\ \gint[][c][cred][a];\gint[][a][token][c];\gint[][c][token][s];\gint[][s][ok][c];\gint[][c][helo][s];\gint[][s][int][c] \\
  \aG_{\qosattr{msg}} = &\ \gint[][c][read][s]; \gint[][s][size][c]; \gint[][c][retr][s]; \gint[][s][msg][c]; \gint[][s][ack][c]
\end{align*}
to check whether the cost of receiving one
email falls below a given threshold.
The validity of $\Phi_1$ is checked by \thetool in less than one second
when using the length of any run matching the g-choreography as a
bound (namely, we fix $\mathsf{k} = 26$).
We can also write the formula 
$\Phi_2 = [\aG_\qosattr{init};{\aG_\qosattr{msg}}^{\grecop}]\
\qosattr{totalCost} \leq \qosattr{emailsRetrieved}$ to check
a relation between the total cost of an execution and the number of
emails retrieved by the client.
In less than one second \thetool produces a counterexample to the
validity of $\Phi_2$ (a run where the client retrieves $0$
emails). This means the cost of such an execution is not $0$, which
hints at the fact that there is a fixed cost for the execution of the
services even if no emails are retrieved.
On the formula
$\Phi_3 = [\aG_\qosattr{init};{\aG_\qosattr{msg}}^{\grecop}]\
\qosattr{totalCost} \leq 1 + \qosattr{emailsRetrieved}$ (which sets at
$1$ USD the fixed cost), \thetool takes 135 seconds to report that no
counterexamples were found with a bound of $\mathsf{k} = 100$.
A stronger requirement can be expressed by putting tighter bounds on
the fixed cost and the cost per email:
$\Phi_4 = [\aG_\qosattr{init};{\aG_\qosattr{msg}}^{\grecop}]\
\qosattr{totalCost} \leq 0.5 + 0.5 \cdot \qosattr{emailsRetrieved}$.
Again, with a bound of $\mathsf{k} = 100$, \thetool takes 137 seconds
to report that no counterexamples were found.

%%% Local Variables:
%%% mode: latex
%%% TeX-master: "main"
%%% End:

\subsection{Model extraction}\label{sec:extraction}
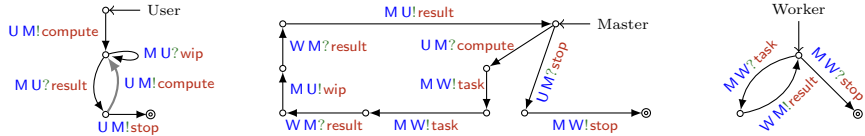
\begin{figure}[h]
    \centering
    \begin{tikzpicture}[node distance=0.5cm, every node/.style={scale=.8,transform shape,font=\scriptsize}, every label/.style={color=blue,fill=yellow!10,scale=.7}]
        % User
        \node[cnode,initial, initial where = right, initial text=User] (u1) {};
        \node (u2) [cnode, below = of u1] {};
        \node (u3) [cnode, below = .7cm of u2] {};
        \node (u4) [cnode, right = of u3, double] {};
        \path[line] (u1) -- node[left]{$\aout[U][M][compute]$} (u2);
        \path[line] (u2) edge[bend left=-35] node[left]{$\ain[M][U][result]$} (u3);
        \path[line] (u3) edge[bend left=-35, color=gray, thick] node[right]{$\aout[U][M][compute]$} (u2);
        \path[line] (u3) -- node[below]{$\aout[U][M][stop]$} (u4);
        \path[line] (u2) edge [loop right] node[right]{$\ain[M][U][wip]$} (u2);
		\end{tikzpicture}
		\qquad
		% Master
		\begin{tikzpicture}[node distance=0.5cm, every node/.style={scale=.8,transform shape,font=\scriptsize}, every label/.style={color=blue,fill=yellow!10,scale=.7}]
		  \node (m7) [cnode] {};
        \node (m6) [cnode, below = of m7] {};
        \node (m5) [cnode, below = of m6] {};
        \node (m4) [cnode, right = 1cm of m5] {};
        \node (m3) [cnode, right = 1.5 of m4] {};
        \node (m2) [cnode, above = of m3] {};
        \node (m1) [cnode, initial, initial where = right, initial text = Master, right = 3.5cm of m7] {};
        \node (m8) [cnode, right = 2cm of m4] {};
        \node (m9) [cnode, accepting, right = 1.5cm of m8] {};
        \path[line] (m1) -- node[left]{$\ain[U][M][compute]$} (m2);
        \path[line] (m2) -- node[above left = -.1cm]{$\aout[M][W][task]$} (m3);
        \path[line] (m3) -- node[below]{$\aout[M][W][task]$}  (m4);
        \path[line] (m4) -- node[below]{$\ain[W][M][result]$} (m5);
        \path[line] (m5) -- node[right]{$\aout[M][U][wip]$} (m6);
        \path[line] (m6) -- node[right]{$\ain[W][M][result]$} (m7);
        \path[line] (m7) -- node[above]{$\aout[M][U][result]$} (m1);
        \path[line] (m1) -- node[sloped, below]{$\ain[U][M][stop]$} (m8);
        \path[line] (m8) -- node[below]{$\aout[M][W][stop]$} (m9);
		\end{tikzpicture}
		\qquad
		% Worker
		\begin{tikzpicture}[node distance=0.5cm, every node/.style={scale=.8,transform shape,font=\scriptsize}, every label/.style={color=blue,fill=yellow!10,scale=.7}]
        \node[cnode, initial, initial where = above, initial text=Worker] (w1) {};
        \node[cnode, below left = 1cm of w1] (w2) {};
        \node[cnode, below right = 1cm of w1, accepting] (w3) {};
        \path[line] (w1) edge[bend left=-35] node[above, sloped]{$\ain[M][W][task]$} (w2);
        \path[line] (w2) edge[bend left=-35] node[below, sloped]{$\aout[W][M][result]$} (w1);
        \path[line] (w1) -- node[sloped, above]{$\ain[M][W][stop]$} (w3);
        % \foreach \s/\l/\t in {0/\specA/1,1/{}/2}{
        %     \node (A\t) [cnode, right = of A\s, label = $\l$] {};
        % }
        % \node (A3) [cnode, right = of A2, fill = black] {};
        % \foreach \s/\l/\t in {0/\aout[a][b][x]/1,2/\aout[a][b][z2]/3}{
        %     \path[line,sloped] (A\s) edge node[below]{$\l$} (A\t);
        % }
        % \path[line] (A1) edge[bend left=-35] node[below]{$\ain[b][a][y]$} (A2);
        % \path[line] (A2) edge[bend left=-35] node[above]{$\aout[a][b][z1]$} (A1);
		\end{tikzpicture}
    \caption{Communicating system of the case study presented in \cite{imai:tacas22}\label{fig:eval:kmclib}}
\end{figure}

% \noindent \textbf{Model extraction}
We show how to apply \thetool on a model automatically inferred from
the OCaml code of the case study presented in~\cite{imai:tacas22}.
\cref{fig:eval:kmclib} depicts the system inferred
in~\cite{imai:tacas22} where the user requests the master to resolve a
computational problem (whose nature is inconsequential here).
The master splits the problem into two tasks and sends them to the
worker, which sends back to the master the solutions of each task.
Before collecting the second partial results from the worker, the
master sends a \squo{work-in-progress} message to the user.
Finally, upon reception of the second partial result,
the master combines the solutions and sends the final result to the
user.
The thick gray arrow in~\cref{fig:eval:kmclib} is the only addition we
made to the original case study so to enable the user to decide to
iteratively make new requests or to stop.

The QoS specifications involve price, number of tasks computed, and
allocated memory, respectively denoted with \qosattr{p}, \qosattr{t},
and \qosattr{mem}.
The contraints over these attributes have been manually specified and assigned to the
states of the qCFSMs.
%
% \eMcomm[I don't get this]{
  For instance, we assume each problem instance (requested from the user) to require at most $5$
units of memory and model this by adding a contraint over \qosattr{mem} to
the states where memory is allocated for the
problem intance.
Similarly, we assume the result of the problem to require at most $1$ unit of memory.
Additionally, we assume that the master charges a flat fee of $10$ monetary units once the computation is completed, while the worker's cost varies based on the size of the task.
We check the following $\QL$ properties:
% $$\begin{array}{lcl@{\qquad}lcl}
%   \Phi_1 & \equiv & [\aG] (\qosattr{t} \cdot 6 \leq \qosattr{p} < 12.5)
%    & \Phi_3  & \equiv & [\grec[][\aG]] (1 \leq \qosattr{mem} < 10)
%   \\
%   \Phi_2 & \equiv & [\grec[][\aG]] (\qosattr t \cdot 6 \leq \qosattr p < 12.5)
%   & \Phi_4 & \equiv & (\qosattr p \leq \qosattr t \cdot 12.5)\ \qluntil ([\grec[][\aG]]\ \qluntil p \leq 25) 
% \end{array}$$
\\[5pt]
$\begin{array}{lcl@{\qquad}lcl}
  \Phi_1  &\equiv& [\aG] (\qosattr{t} \cdot 6 \leq \qosattr{p} < 12.5)
  &
  \Phi_3  &\equiv& [\grec[][\aG]] (1 \leq \qosattr{mem} < 10)
  \\
  \Phi_2 &\equiv& [\grec[][\aG]] (\qosattr t \cdot 6 \leq \qosattr p < 12.5)
  &
  \Phi_4 &\equiv& (\qosattr p \leq \qosattr t \cdot 12.5)\ \qluntil ([\grec[][\aG]]\ \qosattr p \leq 25) 
\end{array}$
\\[5pt]
where $\aG$ describes the process of computing one problem instance, starting with
$\gint[][U][compute][M]$ and ending with $\gint[][M][result][U]$
(\cref{appendix:model-extraction} reports the details about the models of this case study.)
Formula $\Phi_1$ uses the necessity modality to express bounds on the price
of the computation of one problem instance. Formulas $\Phi_2$ and $\Phi_3$
use the necessity modality to express bounds on the price and the memory used
after computing any number of problem instances. 
% 
% We run \thetool on these three formulas and obtain a model within a bound \texttt{k} = 18, 
% which is the length of the the runs that first match $\aG$ and then advance to a final configuration. 
% In each case, \thetool is able to find the model in less than 0.4 seconds.
% % 
% By using the \texttt{validity} command we can also use \thetool as a counterexample finding procedure by checking satisfiability of the negation
% of the formula. With a bound \texttt{k} = 32 it takes, respectively, 35, 206 and 208 seconds
% to report that no counterexamples were found for $\Phi_1, \Phi_2$ and $\Phi_3$.
% 
Formula $\Phi_4$ states that
\begin{inparaenum}[(i)]
    \item up to the computation of the first problem instance, the price falls below a bound depending on the number of tasks computed, and
    \item afterwards, the price is always bounded by 25 right after any number of computed problem
    instances.
\end{inparaenum}
% 
% Using a bound \texttt{k} = 18, it takes \thetool 0.7 seconds to find a model for $\Phi_4$.
% If we use \thetool as a counterexample finding procedure, the bound is not enough
% to find a counterexample: it takes 7 seconds to report that no counterexamples
% were found. However, if we increase the bound to \texttt{k} = 32, \thetool finds a counterexample
% in 35 seconds. Which shows that, unlike the previous formulas, $\Phi_4$ is satisfiable but it is not valid.
% 
We applied \thetool on these formulas with bounds that correspond to
unfolding loops once and twice ($\mathsf{k} = 18$ and
$\mathsf{k} = 32$ are, respectively, the lengths of runs where the
master sends the result of one and two problem instances and the user
stops).
The results of the experiments are summarised in
\cref{model-extraction:tab} where times are in seconds.
\begin{table}[t]
\setlength{\belowcaptionskip}{\captionsep}
  \centering
  \def\mt#1{\multicolumn 1 {c|} {#1}}
\scalebox{.9}{  \begin{tabular}{c|c|c|c|c|c|c|c|c}
	 \multicolumn 1 c {} & \multicolumn 4 {c|}{\bf Bound \sf{k} = 18 } & \multicolumn 4 {c} {\bf Bound \sf{k} = 32 }
	 \\\cline{2-9}
	 \multicolumn 1 c {} & \multicolumn 2 {|c|} {satisfiability} & \multicolumn 2 {c|} {validity} & \multicolumn 2 {c|} {satisfiability} & \multicolumn 2 c {validity}
	 \\\cline{2-9}
	 \bf{ Formula } & \mt{\bf Time (s)} & \mt{\bf Result} & \mt{\bf Time (s)} & \mt{\bf Result} & \mt{\bf Time (s)} & \mt{\bf Result} & \mt{\bf Time (s)} & \multicolumn 1 {c|} {\bf Result}
	 \\\hline
	 $\Phi_1$       & .3                & sat           &  1.7               & No \sf{CE}            & .3                & sat            & 34                & No \sf{CE}
	 \\%\hline
	 $\Phi_2$       & .3                & sat           &  1.9               & No \sf{CE}            & .3                & sat            & 185               & No \sf{CE}
	 \\%\hline
	 $\Phi_3$       & .3                & sat           &  1.9               & No \sf{CE}            & .3                & sat            & 186               & No \sf{CE}
	 \\%\hline
	 $\Phi_4$       & .5                & sat           &  6                & No \sf{CE}         & .6                & sat            & 30                & \sf{CE}
	 \\%\hline	 
  \end{tabular}
}  \caption{Results on model extraction case study (\sf{CE} = counter example)}\label{model-extraction:tab}
\end{table}
Noticeably, for satisfiability the results with $\mathsf{k} = 18$
subsume those with $\mathsf{k} = 32$; also, for $\Phi_4$, a bound of
$32$ is needed to find a counterexample, which shows that the formula
is satisfiable but not valid.

%%% Local Variables:
%%% mode: LaTeX
%%% TeX-master: "main"
%%% End:

\subsection{Performance}\label{sec:perf}
%!TEX root = ./main.tex
%
The performance of \thetool depends on the cost of checking if a formula $\Phi$ holds
on any run of the system $S$ of length at most \texttt{k}; this cost
is dominated by the evaluation of the \squo{until}
sub-formulae\footnote{%
  This requires to query the SMT-LIB to solve QoS constraints; we use
  Z3 as a black-box which we cannot control; therfore, its
  computational costs are factored out.
}
$\Phi_1 \qluntil \Phi_2$ which depends on the complexity of $\aG$
and of $\Phi_1$ and $\Phi_2$.
We therefore generate synthetic properties following the pattern
$\Phi \equiv \Phi_1 \qluntil \Phi_2$ and varying the size and
complexity of $\aG$.
Formulas $\Phi_1$ and $\Phi_2$ are created to cover ($i$) the best
case (any run that matches the language of $\aG$ satisfies $\Phi$),
($ii$) the worst case (no run matching the language of $\aG$ satisfies
$\Phi$), and ($iii$) the `average' case (only a single random run that matches
the language of $\aG$ satisfies $\Phi$).

The performance analysis of \thetool was driven by experiments\footnote{
  We used Z3 v4.10.2 and an 8-cores MacBook Pro (Apple M1) with 16GB
  of memory.  }
tailored to address the following questions:
  \begin{description}
  \item[Loop unfolding] How does performance evolve as we increase the
	 number of loop unfoldings in g-choreographies indexing \squo{until} sub-formulae?
  \item[Nested choices] How does average performance evolve as we
	 increase the number of nested choices in g-choreographies indexing
	 \squo{until} sub-formulae?
  \end{description}
% 
% For the applicability evaluation, we consider the communicating system
% of a case study presented in \cite{imai:tacas22}, which was
% automatically extracted from OCaml code. We manually write QoS
% specifications for the system, propose $\QL$ formulas to be checked in
% the model, and verify them using \thetool.

\noindent
\textbf{Loop unfolding}
The experiments are performed on the set of qCFSMs described in
the AWS cloud case study in \cref{sec:aws}.
To help the reader understand the size of the problem, we show 
below
% in \cref{fig:eval:pop-runs} 
the number of runs of this system as a
function of the bound \texttt{k} on the size of the runs. Due to
interleaving of the transitions in the asynchronous communication, the
number of runs grows exponentially when the number of loop unfoldings
in the POP protocol increases.

\begin{wrapfigure}{l}{0.45\textwidth}
  \centering
  \includegraphics[width=0.45\textwidth]{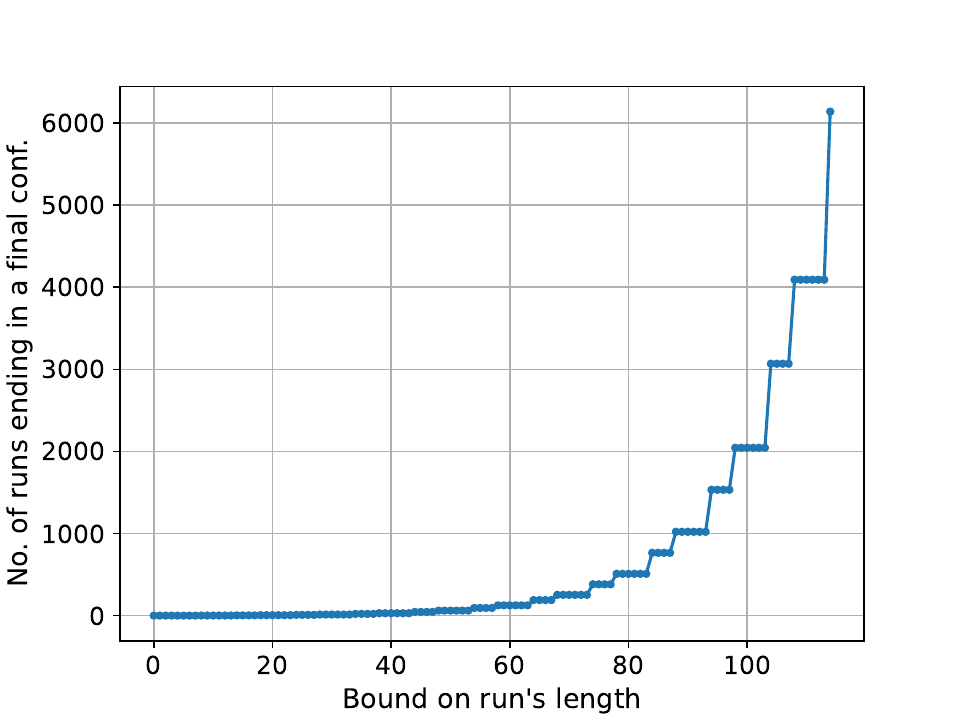}
  % \caption{Size of search space versus the bound on the length of runs (\texttt{k})}
  % \label{fig:eval:pop-runs}
\end{wrapfigure}

\begin{figure}[h]
  \centering
    \begin{subfigure}{0.49\textwidth}
        \centering
        \includegraphics[width=\textwidth]{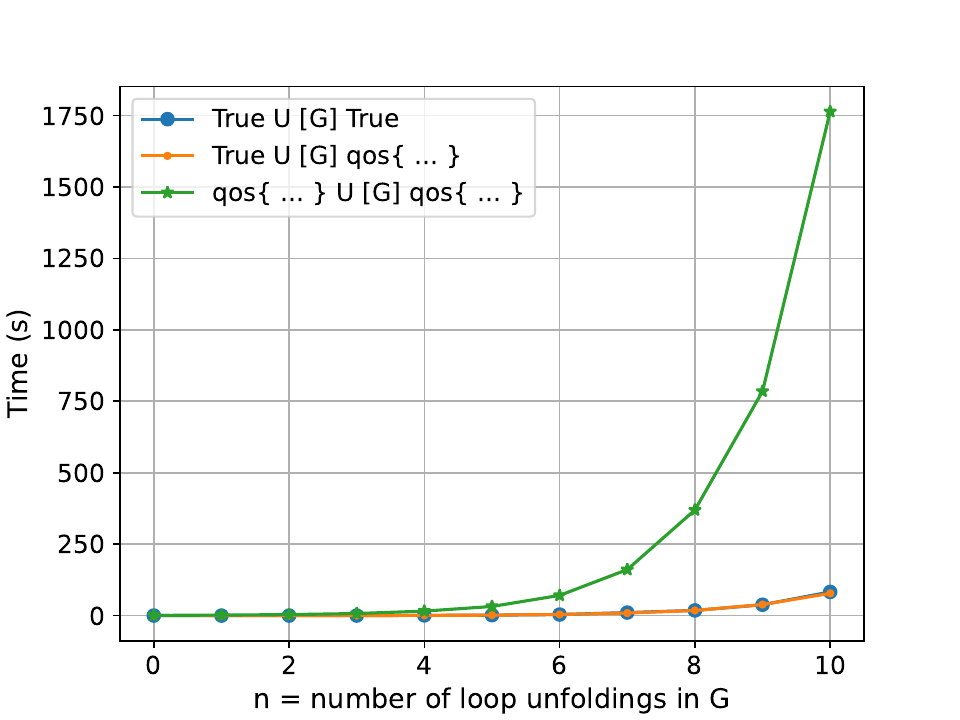}
        \caption{Satisfiable instances}
        \label{fig:eval:sat}
    \end{subfigure}
    \begin{subfigure}{0.49\textwidth}
        \centering
        \includegraphics[width=\textwidth]{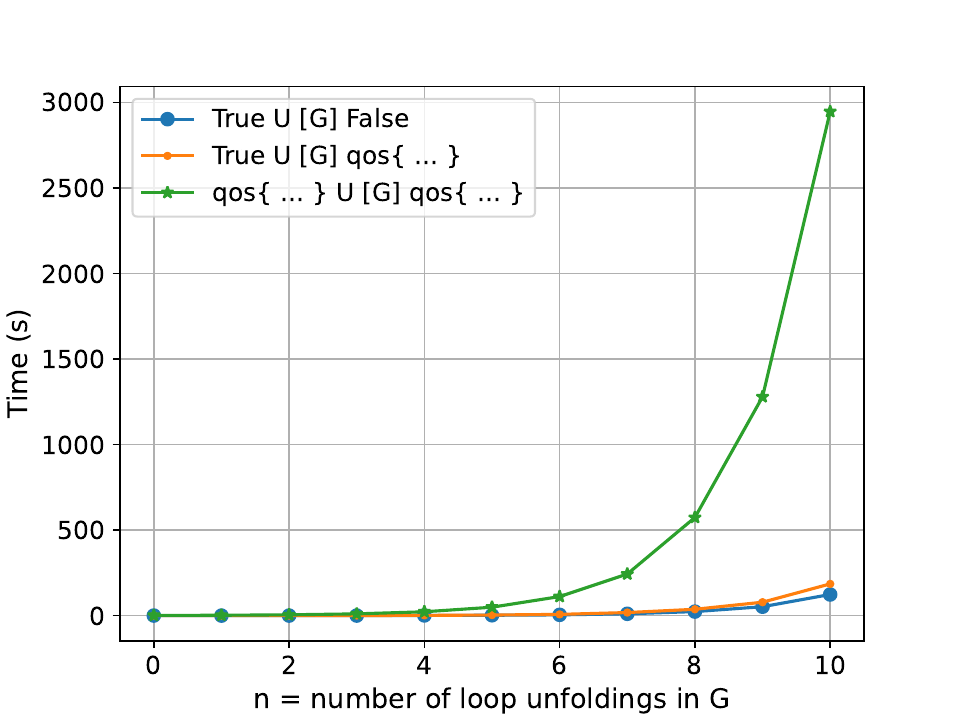}
        \caption{Unsatisfiable instances}
        \label{fig:eval:unsat}
    \end{subfigure}
\setlength{\belowcaptionskip}{-8pt}
    \caption{Execution time to analyse (un)satisfiable \squo{until} formulas}
    \label{fig:eval}
\end{figure}

We synthetically generate six families of $\QL$ formulas with
the shape $\Phi_1 \qluntil \Phi_2$ where
$\aG = \aG_\mathsf{init};{\aG_\mathsf{msg}}^n$, for
$1 \leq n \leq 10$, with $\aG_\mathsf{init}$ 
and $\aG_\mathsf{msg}$ defined as in \cref{sec:aws}.
The first three families of
formulas are constructed to be satisfiable while the last three to be
unsatisfiable.  The unsatisfiable formulas are constructed by
guaranteeing that no run compatible with $\aG$ satisfies $\Phi_2$.
For each formula we execute \thetool with a bound
${\tt k} = 16 + 10n$, which guarantees that the runs of the system
that match $\aG$ are reached.
%
% The graph in \cref{fig:eval:pop-runs} corresponds to values of $n$ ranging 
% from 0 to 10.
%
%
The results are shown in \cref{fig:eval}.
\cref{fig:eval:sat} plots the time it takes for \thetool to find a
model for the three families of satisfiable formulas as a function of
$n$.  The three families differ in how $\Phi_1$ and $\Phi_2$ are
constructed:
\begin{inparaenum}[(i)]
    \item both as atomic truth values,
    \item $\Phi_1$ as an atomic truth value and $\Phi_2$ as a QoS constraint, or
    \item both as QoS constraints.
\end{inparaenum}
\cref{fig:eval:unsat} plots the time \thetool takes to report
that no model was found for the three families of unsatisfiable
formulas as a function of $n$.  The results show that the main source
of computational burden, as the number of loop unfoldings increases,
is the verification of the QoS constraint in $\Phi_1$, the first
operand of the \squo{until} operator. In \cref{fig:eval:sat} and
\cref{fig:eval:unsat}, this is manifested by the green line growing
significantly faster than the other two lines. The explanation for
this is that, due to the semantics of the \squo{until}, the
verification of $\Phi_1$ has to be performed in every prefix of the
run and, when $\Phi_1$ is a QoS constraint, each verification is done
by calling Z3 with a different SMT-LIB query.

\vspace*{1em}

\noindent
\textbf{Nested choices}
% \marginpar{Explain that we will measure the impact of nested choices over synthetic systems.}
To evaluate the performance of \thetool in the presence of nested choices
indexing \squo{until} sub-formulae, we will construct synthetic 
systems by varying the number of nested choices in the g-choreography
of the system.
We consider systems of two participants taking turns in sending a
message to each other; the sender of each turn chooses between two
messages.
Due to the branching nature of this behaviour, the number of runs in a
system grows as $2^n$ where $n$ is %%% an exponential of base 2 that depends on
the number of nested choices (i.e., the number of turns).
We synthetically generate systems with this behaviour by varying
$n$ %%% the number $n$ of nested choices
from 1 to 10.
Remarkably, nested choices correspond to nested conditional statements
and accepted metrics recommend to keep low the nesting level of
conditional statements.
In particular, an accepted upper bound of \emph{cyclomatic complexity}\footnote{
  Cyclomatic complexity~\cite{mccabe:ieeetse-se-2_4} measures the complexity of programs according
  to the number of independent paths represented in the source code.
} is 
15, which corresponds to less than 4 nested conditional statements.
%%% In particular, \emph{cyclomatic complexity} aims to measure the complexity of programs by reflecting the number of independent paths represented in the source code.
    % In~\cite{mccabe:ieeetse-se-2_4} the author draw conclusion of the application of this metric to programs in a production environment from which it can be read:
    % \begin{quote}
    % The particular upper bound that has been used for cyclomatic complexity is 10 which seems like a reasonable, but not magical, upper limit. 
    % \end{quote}
	 %%% A cyclomatic complexity~\cite{mccabe:ieeetse-se-2_4} of up to 10--15, a nowadays accepted upper bound, represents less than 4 conditiona statements, for independent conditions forming a sequence, or 14, for dependant conditions that can not be factored out.%, forming a perfectly balanced binary tree.

To generate these systems, we craft QoS-extended g-choreography in
\texttt{.qosgc} format and then leverage \chorgram's \texttt{G-chor
  projection} to obtain the qCFSMs of the system.
The QoS specifications comprise five QoS attributes and determine 
unique values for them in each final state, enabling the
construction of $\QL$ formulas that are satisfied by only one run of
the system.  In this way, we can use these generated cases to evaluate
the performance of \thetool in finding the only run that satisfies a
formula in a search space of exponential size in $n$.
\begin{figure}[h]
  \begin{subfigure}{0.46\textwidth}
    \begin{lstlisting}[language=sgc,basicstyle=\scriptsize]
$\p[Bob]$ -> $\p[Alice]$: $\msg[m0]$ ; {
  $\p[Alice]$ -> $\p[Bob]$: $\msg[m0]$; $\p[Bob]$ -> $\p[Alice]$: $\msg[leaf1]$
  +
  $\p[Alice]$ -> $\p[Bob]$: $\msg[m1]$; $\p[Bob]$ -> $\p[Alice]$: $\msg[leaf2]$
}

+

$\p[Bob]$ -> $\p[Alice]$: $\msg[m1]$; {
  $\p[Alice]$ -> $\p[Bob]$: $\msg[m0]$; $\p[Bob]$ -> $\p[Alice]$: $\msg[leaf3]$
  +
  $\p[Alice]$ -> $\p[Bob]$: $\msg[m1]$; $\p[Bob]$ -> $\p[Alice]$: $\msg[leaf4]$
}
    \end{lstlisting}%
    \caption{Generated g-choreography for $n = 2$}
    \label{fig:eval:nested-choices-code}
    \end{subfigure}%
    \hfill
    \begin{subfigure}{0.46\textwidth}
        \centering
        \includegraphics[scale=.33]{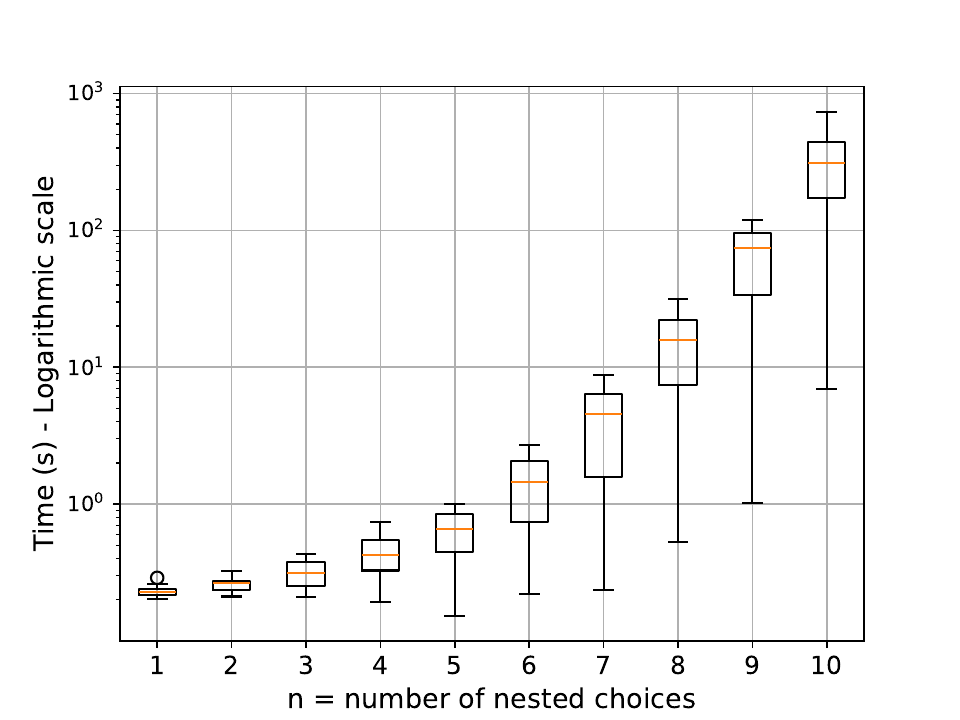}
        \caption{Average execution time versus $n$}
        \label{fig:eval:nested-choices-plot}
    \end{subfigure}%
\setlength{\belowcaptionskip}{-10pt}
\caption{Performance on the `average' case for formulas with nested choices}
\label{fig:eval:nested-choices}
\end{figure}

\noindent
The formula is generated by following the pattern
$\top\ \qluntil\ \psi$, were $\aG$ matches every run of the
system and $\psi$ is a QoS constraint, determining the value of the five QoS
attributes, that is satisfied by only one run.
\cref{fig:eval:nested-choices-code} shows the generated $\aG$ for
two nested choices.
See \cref{appendix:nested-choices} for a detailed view of the 
files used in this case study.
The bound \texttt{k} is set high enough to guarantee that all runs of
the system are reached by the analysis.  For each value of $n$, we
generate 100 different random instances of the $\QL$ formula, where
the only run that satisfies the formula is chosen randomly, and
execute \thetool on each instance.
\cref{fig:eval:nested-choices-plot} shows the results as a boxplot per
number of nested choices.
Remarkably, both the average execution time and its variance grow as
$n$ increases.
This is due to the fact that the difference in time
between the best and worst case scenarios, where the model is found
either in the first or the last enumerated run matching $\aG$,
increases with $n$.
The apparent bias in dispersion towards lower
execution times is just a visual effect due to the logarithmic scale
of the y-axis.

%%% Local Variables:
%%% mode: latex
%%% TeX-master: "main"
%%% End:

\section{Related Work}\label{sec:rw}
%!TEX root = ./main.tex
%
We position \thetool in the category of static analysers of system-level QoS
properties.
There is a vast literature on QoS, spanning a wide range of contexts
and methods~\cite{Aleti2013SoftwareAO,hayyolalam:jnca-110}, QoS for
choreographies~\cite{ivanovic:icsoc12,kattepur:icsoc13}, and
formal models and analysis procedures that have been proposed without
tool supported analysis.

A tool for the automatic analysis of QoS properties appeared
in~\cite{martinezsune:coordination19} where QoS specifications were
expressed as theory presentations over quantitative attributes but
only considering convex polytopes; this restriction is not present in
our language.
Unlike \thetool, the approach in~\cite{martinezsune:coordination19}
relies on \quo{monolithic} specifications of QoS, rendering hard
its application to distributed systems without adding some composition
mechanisms.
We instead assign QoS contracts to states of communicating services
and then aggregate them in order to analyse properties along
executions of the whole system.

Metric functions are used in~\cite{gdgln16} to verify SLAs of
client-server systems via the interactive theorem prover
KeY~\cite{din:cade15}.
We can deal with multiparty system and the analysis of QoS properties
of \thetool is fully automatic.
Other abstract models of QoS such as quantales~\cite{rosenthal90} or
c-semirings~\cite{buscemi:esop07,lluchlafuente:qapl04,denicola:coordination05}
have been proposed.
Process calculi capable of expressing SLAs appeared
in~\cite{buscemi:esop07} and in~\cite{denicola:coordination05} without
a specific analysis technique.
A variant of the $\mu$-calculus equipped with the capability of
expressing QoS properties and an analysis algorithm has been presented
in~\cite{lluchlafuente:qapl04} without an implementation.

Automatic extraction of local QoS contracts from global QoS
specifications is defined and implemented in~\cite{ivanovic:icsoc12};
the paper proposes applications including the use of the derived
contracts for
monitoring %%% for monitoring the behaviour of components, %%% both on-line and off-line,
but no static analysis procedure of the QoS systems' behaviour is
proposed.
On the same basis, monitoring algorithms were presented
in~\cite{kattepur:icsoc13} and contracts are used for run-time
prediction, adaptive composition, or compliance checking.

Probabilistic model checking (PMC)~\cite{kwiatkowska:esec07,baier08}
implemented in PRISM~\cite{kwiatkowska:cav11} features the automated
analysis of quantitative properties.
The main differences with respect to our work are the modelling
language and the properties that can be checked.
First, PMC models are usually expressed as Markov chains while
\thetool does not feature probabilistic information.
Second, RCFs are more expressive than the reward functions adopted
in~\cite{kwiatkowska:esec07} since they allow to express first order
formulae over QoS attributes.
For example, the QoS specifications shown in \cref{sec:aws} cannot be expressed with PRISM's
reward functions.
Finally, while in PMC properties are expressed as temporal formulae
over bounds on the expected cumulative value of a reward, \thetool can
verify dynamic temporal formulae where atoms are first order formulae
over QoS attributes and temporal modalities are indexed with
g-choreographies. Our setting leads to the computation of an
aggregation function that collects QoS specifications of states along
a run, which is not the case in PMC.
%Conversely, our models are not probabilistic and thus we cannot express probabilistic properties. A combination of the two approaches is an interesting direction for future work.
% 
%Similarly, the verification of real-time systems modeled as timed automata~\cite{alur:tcs-126_2}, supported by tools such as UPPAAL~\cite{Larsen1997UppaalIA}, does not subsume our approach. Our QoS constraints attached to states do not act as control conditions but act as specifications that the implementation of each service must satisfy. Moreover, these specifications predicate over attributes of (potentially) diverse nature, while timed automata conditions predicate over clock variables representing the passage of time.
Timed automata~\cite{alur:tcs-126_2} are used in
UPPAAL~\cite{Larsen1997UppaalIA} to verify real-time systems.
Our QoS specifications can predicate about time but, unlike in UPPAAL,
the behaviour of systems is independent of it.
%The role of these QoS specifications in the semantics of our logic cannot be emulated by timed automata. Conversely, our models are not timed and thus we cannot express real-time properties. Adapting our approach to timed systems could be a promising direction for future work.
It is therefore not straightforward to compare \thetool with tools like PMC or UPPAAL as they are designed for different purposes.  Extending \thetool with time and probabilities is indeed an intriguing endeavour.

%%% Local Variables:
%%% mode: latex
%%% TeX-master: "main"
%%% End:

\section{Conclusions and future work}\label{sec:conclu}
% !TEX root = ./main.tex
We presented \thetool, a tool to verify QoS properties of
message-passing systems. We build a bounded model checker upon the
dynamic logic and semi-decision procedure recently
presented in~\cite{lopezpombo:ictac23} which rely on choreographic models.
To our best knowledge, \thetool is the first tool to support the
static analysis of QoS for choreographic models of message-passing
systems.
The satisfiability of QoS constraints in atomic formulas is delegated
to the SMT solver Z3 while \chorgram is used to handle the
choreographic models and their semantics.
Notably, \thetool can handle any quality attribute that takes values 
in the real numbers (if it is equipped with an
appropriate aggregation operator), making it highly versatile.
Experiments to evaluate the applicability of our approach were
conducted over case studies based on the AWS cloud and on models
automatically extracted from code.
Experiments to evaluate the scalability of our approach
were conducted over synthetically generated models and properties.

% \aGcomm[
%   Clarify that these optimisations describe possible improvements concerning efficiency and not affecting the functionalities of MoCheQoS.
%   We left these optimisations for future work because in our experiments we did not find any limitation of the usability of MoCheQoS due to the lack of these optimisations.
% ]{
Our experiments demonstrate the effectiveness of \thetool.
Nevertheless, there is room for improvement.
We are considering abstract semantics where runs are partitioned in
equivalence classes so that we have to check only representative runs
of such classes in order to tackle the computational blow up due to
asynchronous communications as discussed in \cref{sec:eval}.
% %
% Also, we plan to reduce the execution time of the check that runs
% belong to the language of a g-choreography by replacing the pomset
% semantics implemented in \chorgram with one based on event structures.
% %
% In fact, event structures, akin tries, represent runs with common
% prefixes with a tree-like structures (where sibling nodes are in
% conflict). In this way we avoid duplicating the checks on common prefixes
% as currently done with the use of pomsets.
%
% We however argue that, in practice, this is a minor issue since we
% do not expect to have high number of nested choices in realistic
% g-choreographies.

In scope of future work is also the definition of a domain-specific
language to ease the modelling phase. For instance, such language
could feature data types to express non-cumulative attributes (as
those used in \cref{sec:aws}).

\bibliographystyle{splncs}
\bibliography{bibdatabase,morebiblio}

\newpage

\appendix

\section{SLA in the amazon cloud}\label{appendix:aws}
In \cref{appendix:fig:loop-unfoldings-cfsms} and \cref{appendix:fig:loop-unfoldings-qosfsa}
we show, respectively, the qCFSMs and the QoS specifications for the three-party POP protocol 
used in the case study based on AWS (c.f. \cref{sec:aws}) and loop unfoldings experiments (c.f. \cref{sec:perf}). 
In \cref{appendix:fig:loop-unfoldings-lts} we show the transition system.

\vfill
\def\datap{\Gamma_\mathrm{data'}}
\def\initS{\Gamma_\mathrm{init_{\p[s]}}}
\def\initA{\Gamma_\mathrm{init_{\p[a]}}}
\def\email{\Gamma_\mathrm{email}}
\def\compA{\Gamma_\mathrm{comp_{\p[a]}}}
\def\compC{\Gamma_\mathrm{comp_{\p[c]}}}
\def\compS{\Gamma_\mathrm{comp_{\p[s]}}}
\begin{figure}
\begin{align*}
    \aCM_{\p[a]} = &
    \begin{tikzpicture}[node distance=1cm and 1.3cm,transform shape, every label/.style={color=blue,fill=yellow!10,scale=.75}]
       \node[cnode, initial, initial text = {}, label={below:$\initA$}] (0) {};
       \node (1) [cnode, right = of 0, label={below left:$\compA$}] {};
       \node (2) [cnode, above right = of 1, label={above:$\datap$}] {};
       \node (3) [cnode, right = of 1] {};
        \path[line,sloped] (0) edge node[above]{$\ain[a][c][cred]$} (1);
        \path[line,sloped] (1) edge node[above]{$\aout[c][a][token]$} (2);
        \path[line,sloped] (1) edge node[below]{$\aout[c][a][error]$} (3);
    %    \node (-4) [cnode, above = of -2] {};
    %    \path[line] (-4) edge node[above]{$\ain[a][c][token]$} (-3);
    %    \foreach \s/\l/\t in {-1//0,0/\comp/1,1//2,2/\comp/3,3//4} {
    %       \node (\t) [cnode, right = of \s, label={above:$\l$}] {};
    %   }
    %   \node[cnode, above = of -1] (-3) {};
    %       \node (5) [cnode, above = of 4, label={above right:$\comp$}] {};
    %       \node (6) [cnode, left = of 5, label={$\data$}] {};
    %       \path[line] (-3) edge node[left]{$\aout[c][s][token]$} (-1);
    %    \foreach \s/\l/\t in {-1/\ain[s][c][ok]/0,0/\aout[c][s][helo]/1,1/\ain[s][c][int]/2,2/\aout[c][s][read]/3,3/\ain[s][c][size]/4} {
    %       \path[line,sloped] (\s) edge node[below]{$\l$} (\t);
    %   }
    %    \path[line] (4) edge node[right]{$\aout[c][s][retr]$} (5);
    %    \path[line] (5) edge node[above, pos = 0.4]{$\ain[s][c][msg]$} (6);
    %    \node[cnode, above = of 0, label={95:$$}] (7) {};
    %    \node[cnode, accepting, right = of 7, label={10:$$}] (8) {};
    %    %
    %    \path[line, sloped] (6) edge node[pos=0.35, above]{$\aout[c][s][ack]$} (2);
    %    \path[line] (0) edge node[left]{$\aout[c][s][quit]$} (7);
    %    \path[line] (2) edge[sloped] node[below,near end]{$\aout[c][s][quit]$} (7);
    %    \path[line] (4) edge[sloped] node[pos = 0.6, above = -0.05]{$\aout[c][s][quit]$} (7);
    %    \path[line] (7) edge node[above]{$\ain[s][c][bye]$} (8);
    %    %
    %    \node (-5) [cnode, below = of -1] {};
    %    \path[line] (-1) edge node[left]{$\ain[c][s][fail]$} (-5);
    \end{tikzpicture}\\
% \end{align*}
% \begin{align*}
    \aCM_{\p[c]} = &
    \begin{tikzpicture}[node distance=1cm and 1.3cm,transform shape, every label/.style={color=blue,fill=yellow!10,scale=.75}]
       \node[cnode, initial, initial text = {}, label={below:$\compC$}] (-2) {};
       \node (-1) [cnode, right = of -2, label={above right:$\datap$}] {};
       \node (-4) [cnode, above = of -2, label={left:$\datap$}] {};
       \path[line] (-2) edge node[left]{$\aout[c][a][cred]$} (-4);
       \path[line] (-4) edge node[above]{$\ain[a][c][token]$} (-3);
       \foreach \s/\l/\t in {-1//0,0//1,1/\compC/2,2//3,3//4} {
          \node (\t) [cnode, right = of \s, label={[above=0.1em]:$\l$}] {};
          }
      \node[cnode, above = of -1] (-3) {};
          \node (5) [cnode, above = of 4, label={above right:$\compC$}] {};
          \node (6) [cnode, left = of 5, label={$\email$}] {};
          \path[line] (-3) edge node[left]{$\aout[c][s][token]$} (-1);
       \foreach \s/\l/\t in {-1/\ain[s][c][ok]/0,0/\aout[c][s][helo]/1,1/\ain[s][c][int]/2,2/\aout[c][s][read]/3,3/\ain[s][c][size]/4} {
          \path[line,sloped] (\s) edge node[below]{$\l$} (\t);
      }
       \path[line] (4) edge node[right]{$\aout[c][s][retr]$} (5);
       \path[line] (5) edge node[above, pos = 0.4]{$\ain[s][c][msg]$} (6);
       \node[cnode, above = of 0, label={95:$$}] (7) {};
       \node[cnode, accepting, right = of 7, label={10:$$}] (8) {};
       \path[line, sloped] (6) edge node[pos=0.35, above]{$\aout[c][s][ack]$} (2);
       \path[line] (0) edge node[left]{$\aout[c][s][quit]$} (7);
       \path[line] (2) edge[sloped] node[below,near end]{$\aout[c][s][quit]$} (7);
       \path[line] (4) edge[sloped] node[pos = 0.6, above = -0.05]{$\aout[c][s][quit]$} (7);
       \path[line] (7) edge node[above]{$\ain[s][c][bye]$} (8);
       \node (-5) [cnode, below = of -1] {};
       \path[line] (-1) edge node[left]{$\ain[c][s][fail]$} (-5);
    \end{tikzpicture}\\
% \end{align*}
% \begin{align*}
    \aCM_{\p[s]} = &
    \begin{tikzpicture}[node distance=1cm and 1.3cm,transform shape, every label/.style={color=blue,fill=yellow!10,scale=.75}]
       \node[cnode, initial, initial text = {}, label={$\initS$}] (-2) {};
       \node (-1) [cnode, right = of -2, label={above right:$\compS$}] {};
       \foreach \s/\l/\t in {-1//0,0/\compS/1,1//2,2/\compS/3,3//4} {
          \node (\t) [cnode, right = of \s, label={above:$\l$}] {};
      }
      \node[cnode, above = of -1] (-3) {};
          \node (5) [cnode, above = of 4, label={above right:$\compS$}] {};
          \node (6) [cnode, left = of 5, label={$\data$}] {};
       \foreach \s/\l/\t in {-2/\ain[c][s][token]/-1,-1/\aout[s][c][ok]/0,0/\ain[c][s][helo]/1,1/\aout[s][c][int]/2,2/\ain[c][s][read]/3,3/\aout[s][c][size]/4} {
          \path[line,sloped] (\s) edge node[below]{$\l$} (\t);
      }
       \path[line] (4) edge node[right]{$\ain[c][s][retr]$} (5);
       \path[line] (5) edge node[above, pos = 0.4]{$\aout[s][c][msg]$} (6);
       \node[cnode, above = of 0, label={95:$$}] (7) {};
       \node[cnode, accepting, right = of 7, label={10:$$}] (8) {};
       \path[line, sloped] (6) edge node[pos=0.35, above]{$\ain[c][s][ack]$} (2);
       \path[line] (0) edge node[left]{$\ain[c][s][quit]$} (7);
       \path[line] (2) edge[sloped] node[below,near end]{$\ain[c][s][quit]$} (7);
       \path[line] (4) edge[sloped] node[pos = 0.6, above = -0.05]{$\ain[c][s][quit]$} (7);
       \path[line] (7) edge node[above]{$\aout[s][c][bye]$} (8);
       \path[line] (-1) edge node[left]{$\aout[c][s][fail]$} (-3);
    \end{tikzpicture}
\end{align*}
\caption{qCFSMs for the POP protocol.}
\label{appendix:fig:loop-unfoldings-cfsms}
\end{figure}
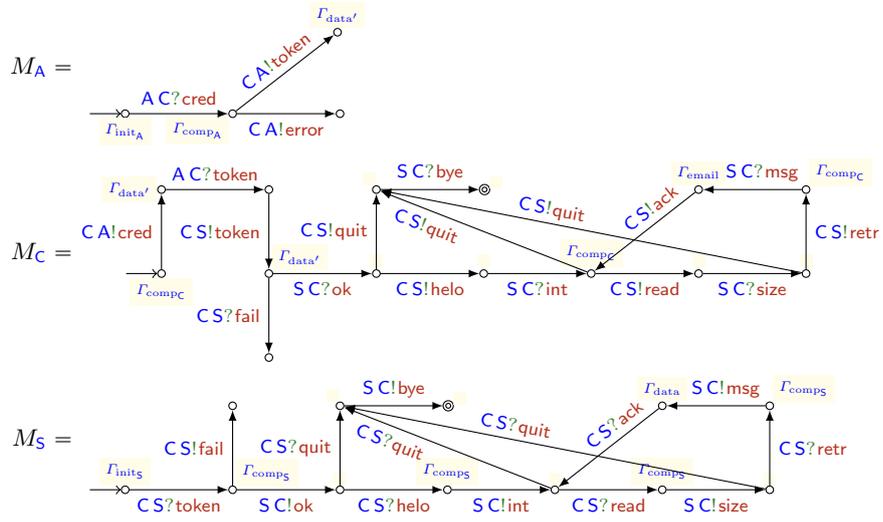

% \vspace*{1cm}

\begin{figure}
% {\fontsize{5}{6}\selectfont
\begin{align*}
\compA = \{\ & 0.5 < \qosattr{execTime} < 3 \land \qosattr{usersAuth} = 1 \ \}\\
\compC = \{\ & 0.5 < \qosattr{execTime} < 3 \ \}\\
\compS = \{\ & 0.5 < \qosattr{execTime} < 3, \qosattr{execTimeServer} = \qosattr{execTime} \ \}\\
\data = \{\ & 10< \qosattr{dataTransferredOut} < 500 \ \}\\
\datap = \{\ & \qosattr{dataTransferredOut} = 2 \ \}\\
\email = \{\ & \qosattr{priceEmails} = 0.0001 \land \qosattr{emailsRetrieved} = 1 \ \}\\
\initA = \{\ & \qosattr{ratePerUserAuth} = 0.001 \ \}\\
% \end{align*}
% \begin{align*}
  \initS = \{\ & \qosattr{instanceType} = 1 \implies (\ \qosattr{hrRateCompute} = 0.0042 \land \qosattr{CPUs} = 2 \\ 
  & \land \qosattr{memoryCapacity} = 0.5 \land \qosattr{networkPerformance} \leq 5\ ), \\
  & \qosattr{instanceType} = 2 \implies (\ \qosattr{hrRateCompute} = 0.0084 \land \qosattr{CPUs} = 2 \\
    & \land \qosattr{memoryCapacity} = 1 \land \qosattr{networkPerformance} \leq 5\ ), \\
    & \qosattr{instanceType} = 3 \implies (\ \qosattr{hrRateCompute} = 0.1344 \land \qosattr{CPUs} = 4 \\
    & \land \qosattr{memoryCapacity} = 16 \land \qosattr{networkPerformance} \leq 5\ ), \\
    & \qosattr{hrRateServerSoftware} = 0.1 \land \qosattr{transferGBRate} = 0.09 \land \qosattr{instanceType} = 1\  
  \}
\end{align*}
% }
\caption{QoS specifications for the POP protocol.}
\label{appendix:fig:loop-unfoldings-qosfsa}
\end{figure}

% \begin{figure}[H]
%     \centering
%     \lstinputlisting[style=qosfsa]{appendix/loop-unfoldings/POP.qosfsa}
%     \caption{\texttt{.qosfsa} file for the loop unfoldings case study.}
%     \label{appendix:fig:loop-unfoldings-qosfsa}
% \end{figure}

% \begin{figure}[H]
%     \centering
%     \includegraphics[width=0.8\textwidth]{appendix/loop-unfoldings/POP-machines.pdf}
%     \caption{CFSMs of the loop unfoldings case study.}
%     \label{appendix:fig:loop-unfoldings-cfsms}
% \end{figure}

\begin{figure}
    \centering
    \includegraphics[scale=0.2]{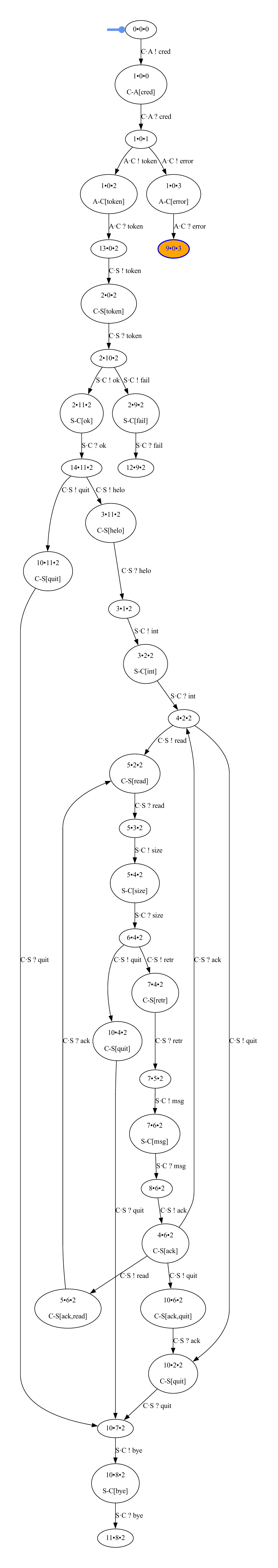}
    \caption{Transition system of the loop unfoldings case study.}
    \label{appendix:fig:loop-unfoldings-lts}
\end{figure}

\clearpage

\newpage

\section{Nested choices}\label{appendix:nested-choices}
In \cref{appendix:fig:nested-choices-qosgc} we show the complete \texttt{.qosgc} file for the case study with number of nested choices = 2. 
This file format is the extension of the \texttt{.gc} file format,
	 which is a sugared version of regular
	 expressions of g-choreographies where
	 $\aG_1 \gchoop \cdots \gchoop \aG_n$ and $\grec[]$ are
	 respectively written
	 $\gselkw\ \p\ {\bracketColor{\{}}\aG_1 + \ldots +
	 \aG_n{\bracketColor{\}}}$ and
	 $\grepkw\ \p\ \bracketColor{\{}\aG\bracketColor{\}}$ where \p\ is
	 the participant enabled to select one of the branches or to stop
	 the iteration.
The \texttt{.qosgc} extension, as discussed in \cref{sec:tool}, contains QoS specifications directly attached to one of the four local states involved in an interaction: \texttt{sqos} denotes the state of the sender before the ouput action, \texttt{rqos} the state of the receiver before the input action, \texttt{sqos'} the state of the sender after the output action, and \texttt{rqos'} the state of the receiver after the input action.
The \texttt{qos} environment that appears in line \texttt{25} is used to declare the QoS attributes and their aggregation operators.
In \cref{appendix:fig:nested-choices-ql} we show the \texttt{.ql} file for one of the formulas used in the case study with number of nested choices = 2; the value used in the QoS constraint of line \texttt{22} was randomly chosen and guaranteed to be satisfied by only one run of the system (see line \texttt{21} in \cref{appendix:fig:nested-choices-qosgc}).
In \cref{appendix:fig:nested-choices-cfsms,appendix:fig:nested-choices-ts} we show the CFSMs and transition system, respectively, for the case study in \cref{appendix:fig:nested-choices-qosgc}.

\vfill

\begin{figure}[H]
    \centering
\begin{lstlisting}[style=ql]
sel Bob {
    Bob -> Alice: m0 ;
    sel Alice {
        Alice -> Bob: m0 ;
        Bob -> Alice: leaf1
            { sqos: (= x1 1) (= x2 1) (= x3 1) (= x4 1) (= x5 1) }
        +
        Alice -> Bob: m1 ;
        Bob -> Alice: leaf2
            { sqos: (= x1 2) (= x2 2) (= x3 2) (= x4 2) (= x5 2) }
    }
    +
    Bob -> Alice: m1 ;
    sel Alice {
        Alice -> Bob: m0 ;
        Bob -> Alice: leaf3
            { sqos: (= x1 3) (= x2 3) (= x3 3) (= x4 3) (= x5 3) }
        +
        Alice -> Bob: m1 ;
        Bob -> Alice: leaf4
            { sqos: (= x1 4) (= x2 4) (= x3 4) (= x4 4) (= x5 4) }
    }
}

 qos {x1:+,x2:+,x3:+,x4:+,x5:+}
\end{lstlisting}
    \caption{\texttt{.qosgc} file for the case study with nested choices = 2.}
    \label{appendix:fig:nested-choices-qosgc}
\end{figure}

\begin{figure}[H]
    \centering
\begin{lstlisting}[style=ql]
True U [
sel Bob {
    Bob -> Alice: m0 ;
    sel Alice {
        Alice -> Bob: m0 ;
        Bob -> Alice: leaf1
        +
        Alice -> Bob: m1 ;
        Bob -> Alice: leaf2
    }
    +
    Bob -> Alice: m1 ;
    sel Alice {
        Alice -> Bob: m0 ;
        Bob -> Alice: leaf3
        +
        Alice -> Bob: m1 ;
        Bob -> Alice: leaf4
    }
}
]
qos{(= x1 4) (= x2 4) (= x3 4) (= x4 4) (= x5 4)}
\end{lstlisting}
    \caption{\texttt{.ql} for one of the formulas in the case study with nested choices = 2.}
    \label{appendix:fig:nested-choices-ql}
\end{figure}

\begin{figure}[H]
    \centering
    \includegraphics[width=\textwidth]{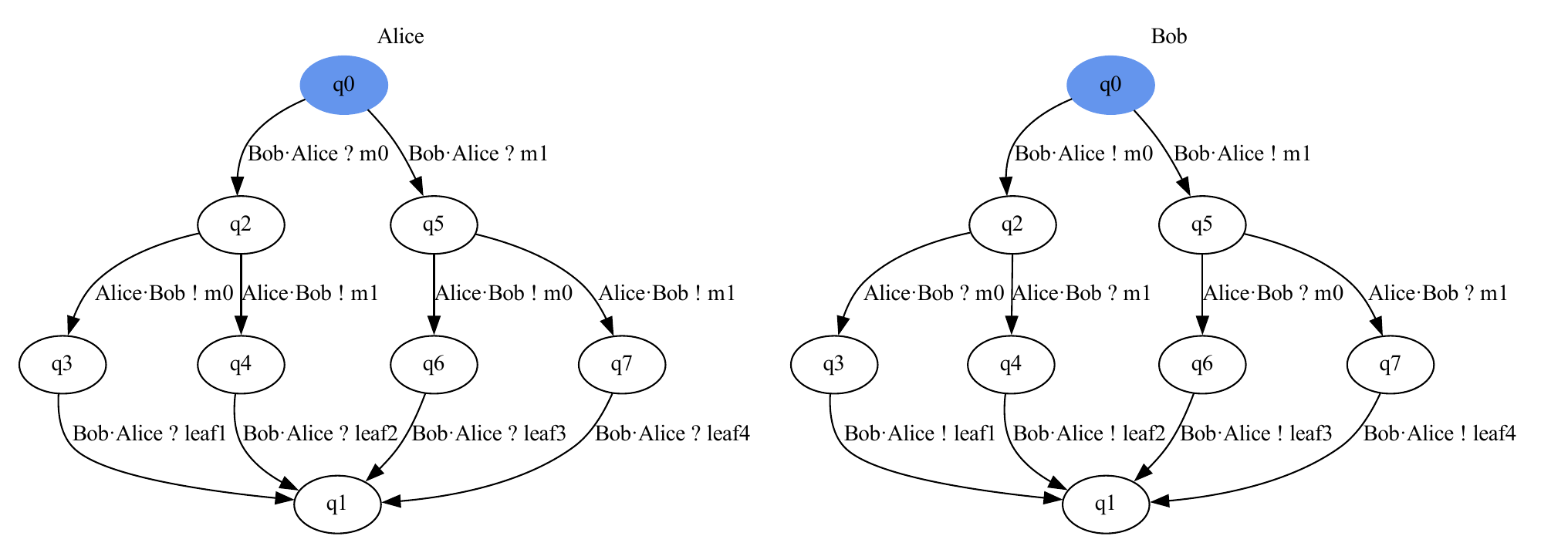}
    \caption{CFSMs of the case study with nested choices = 2.}
    \label{appendix:fig:nested-choices-cfsms}
\end{figure}

\begin{figure}[H]
    \centering
    \includegraphics[width=\textwidth]{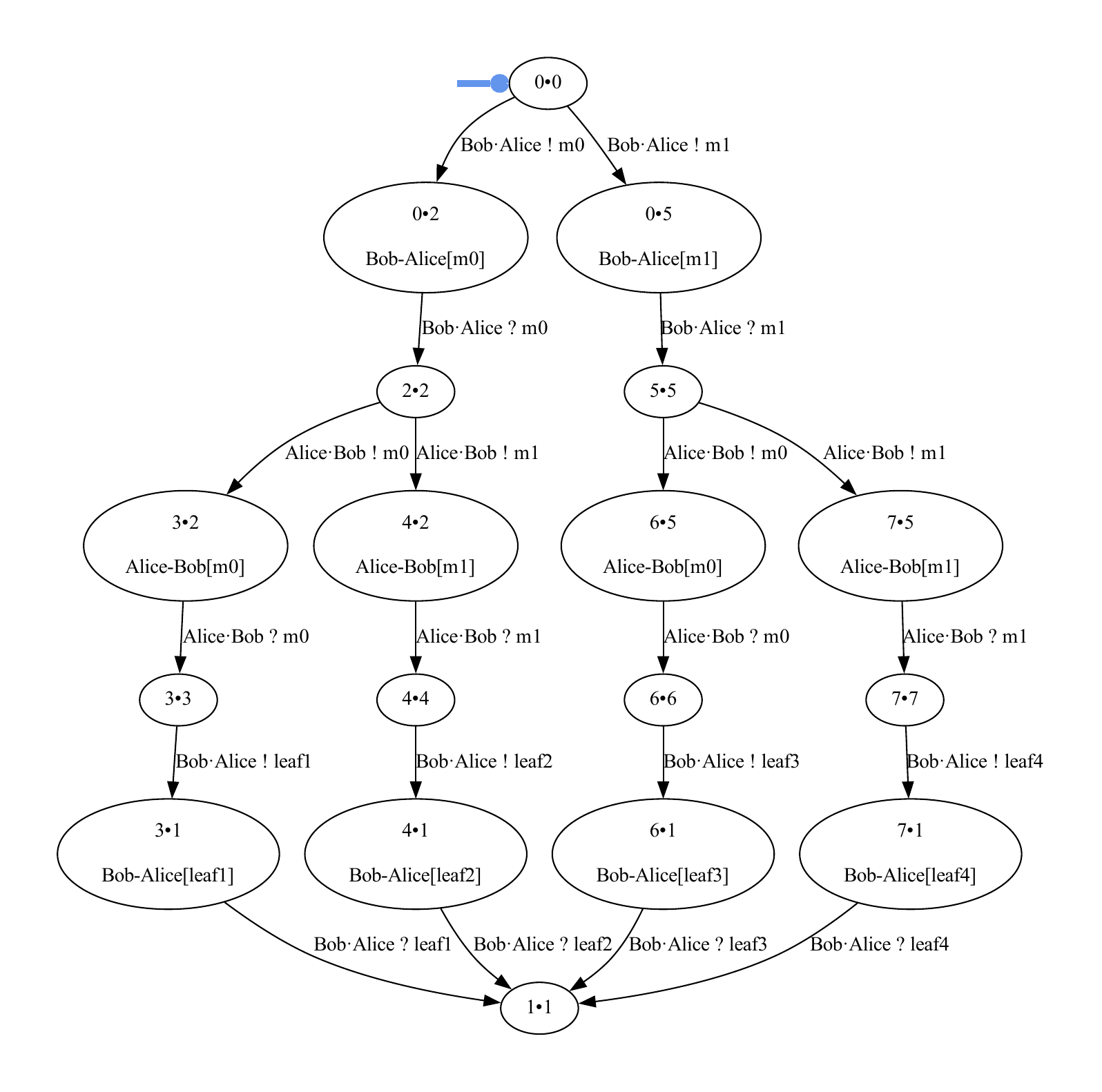}
    \caption{Transition system of the case study with nested choices = 2.}
    \label{appendix:fig:nested-choices-ts}
\end{figure}

\newpage

\section{Model extraction}\label{appendix:model-extraction}
In \cref{appendix:fig:model-extraction-qosfsa} we show the complete \texttt{.qosfsa} file for the model extraction case study; the \texttt{fsa} section of the file is omitted for brevity.
In \cref{appendix:fig:model-extraction-cfsms} and \cref{appendix:fig:model-extraction-ts} we show the CFSMs and transition system, respectively.
In \cref{appendix:fig:model-extraction-ql} we show the \texttt{.ql} file of property $\Phi_4$ used in the case study. The g-choreographies in $\Phi_4$ employ the parallel composition operator ($\gparop$) which is included in the original presentation of the algorithm~\cite{lopezpombo:ictac23} and whose semantics is formally defined in~\cite{tuosto:jlamp-95}.

\vfill

\begin{figure}[H]
    \centering
\begin{lstlisting}[style=qosfsa]
fsa{
    ...
}

qos_attributes{
    p:+,
    t:+,
    pmem:max
}

qos_specifications{
    U@0: (and (< 0 pmem) (< pmem 5) (= t 0) (= p 0)),
    U@2: (and (= pmem 1) (= t 0) (= p 0)),

    M@3: (and (< 0 pmem) (< pmem 10) (= t 0) (= p 0)),
    M@6: (and (= pmem 1) (= t 0) (= p 0)),
    M@8: (and (= pmem 1) (= p 10) (= t 0)),

    W@2: (and (< 0 pmem) (< pmem (+ (/ 5 2) 1)) (= t 1) (=> (<= (- pmem 1) 1) (= p 1)) (=> (> (- pmem 1) 1) (= p (* 3 (- pmem 1)))))
}

final_states{
    U: [3],
    M: [2],
    W: [1]
}
\end{lstlisting}
    \caption{\texttt{.qosfsa} file for the model extraction case study.}
    \label{appendix:fig:model-extraction-qosfsa}
\end{figure}

\begin{figure}[H]
    \centering
    \includegraphics[width=\textwidth]{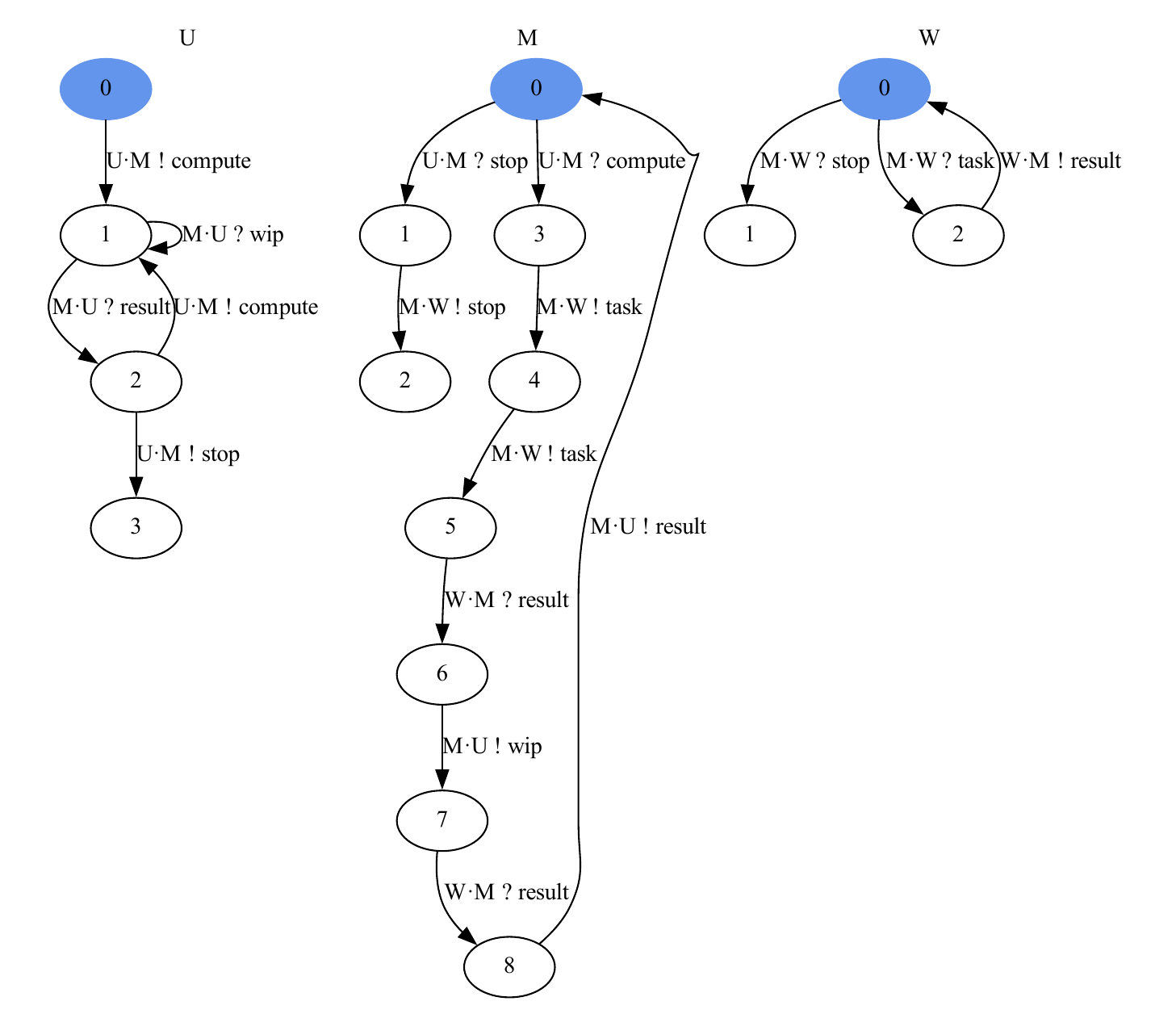}
    \caption{CFSMs for the model extraction case study.}
    \label{appendix:fig:model-extraction-cfsms}
\end{figure}

\begin{figure}[H]
    \centering
    \includegraphics[scale=0.25]{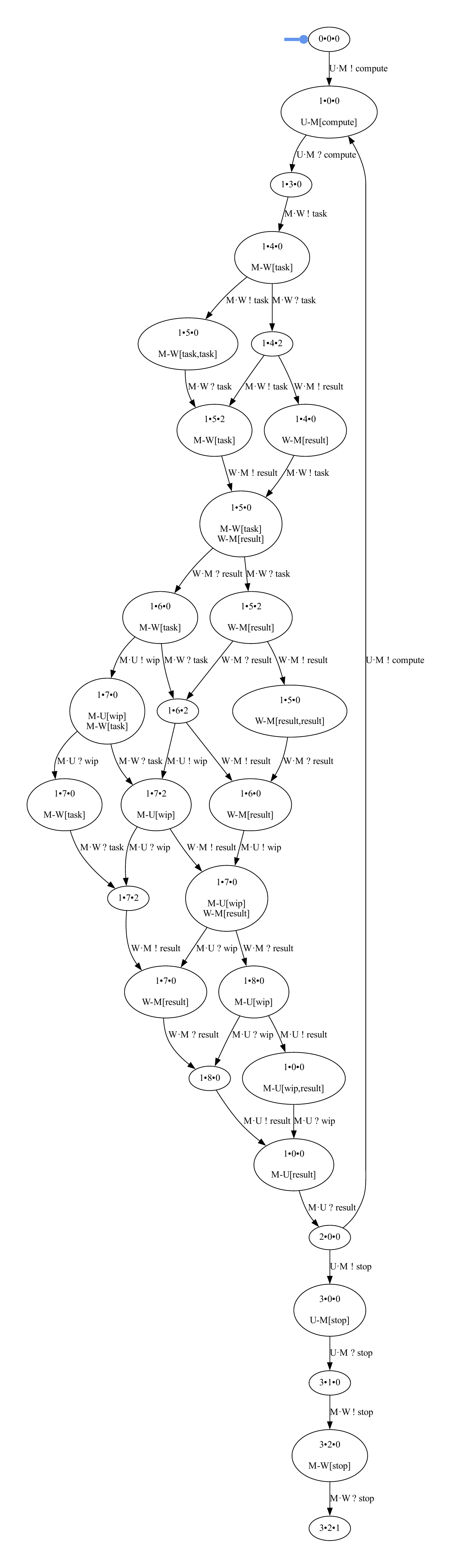}
    \caption{Transition system of the model extraction case study.}
    \label{appendix:fig:model-extraction-ts}
\end{figure}

\begin{figure}[H]
    \centering
\begin{lstlisting}[style=ql]
qos{(<= p (* 12.5 t))} U 
    [
        U -> M : compute;
        M -> W : task;
        {
            M -> W : task
            |
            W -> M : result
        };
        {
            M -> U : wip
            |
            W -> M : result
        };
        M -> U : result
    ]
    (Not (True U
        [
            repeat {
                U -> M : compute;
                M -> W : task;
                {
                    M -> W : task
                    |
                    W -> M : result
                };
                {
                    M -> U : wip
                    |
                    W -> M : result
                };
                M -> U : result
            }
        ]   
        (Not qos{(<= p (* 12.5 2))})
    ))
\end{lstlisting}
    \caption{\texttt{.ql} file of property $\Phi_4$ for the model extraction case study.}
    \label{appendix:fig:model-extraction-ql}
\end{figure}

\end{document}

%%% Local Variables:
%%% mode: latex
%%% TeX-master: t
%%% End: